\documentclass{aa}  

\usepackage{graphicx}
\usepackage{txfonts}
\newcommand{\fe}[1]{Fe\,{\sc ii}~$\lambda#1$}
\newcommand{\oi}[1]{O\,{\sc i}~$\lambda#1$}
\newcommand{\kms}{km\,s$^{-1}$}

\newcommand{\of}{[O\,{\sc i}]~$\lambda6300$}

\defcitealias{HegerWoosley2010}{HW10}

\begin{document}

   \title{A survey of extremely metal-poor gas at cosmic noon}

   \subtitle{Evidence of elevated [O/Fe]}

   \author{Louise Welsh\inst{1, 2, 3} \fnmsep\thanks{Email: louise.welsh@inaf.it}
          \and
          Ryan Cooke\inst{4}
          \and
          Michele Fumagalli\inst{2, 1}
          \and
          Max Pettini\inst{5}
          \and 
          Gwen C. Rudie\inst{6}
          }

\institute{\inst{1} INAF - Osservatorio Astronomico di Trieste, via G. B. Tiepolo 11, I-34143 Trieste, Italy \\
\inst{2} Dipartimento di Fisica G. Occhialini, Universit\`a degli Studi di Milano Bicocca, Piazza della Scienza 3, I-20126 Milano, Italy \\
\inst{3} IFPU - Institute for Fundamental Physics of the Universe, via Beirut 2, I-34151 Trieste, Italy \\
\inst{4} Centre for Extragalactic Astronomy, Durham University, South Road, Durham DH1 3LE, UK \\
\inst{5} Institute of Astronomy, University of Cambridge, Madingley Road, Cambridge CB3 0HA, UK \\
\inst{6} The Observatories of the Carnegie Institution for Sciences, 813 Santa Barbara Street, Pasadena, CA, USA}

   \date{Received 19 June 2024; accepted 11 September 2024}

  \abstract
  {}
   {We aim to study the high-precision chemical abundances of metal-poor gas clouds at cosmic noon ($2 < z < 4$) and investigate the associated enrichment histories. }
  {We analyse the abundances of four newly discovered metal-poor gas clouds utilising observations conducted with Keck/HIRES and VLT/UVES. These systems are classified as very metal-poor (VMP), with [Fe/H]~$<-2.57$, and one system qualifies as an extremely metal-poor (EMP) Damped Lyman-$\alpha$ (DLA) system with [Fe/H]~$=-3.13\pm0.06$. In combination with new high-resolution data of two previously known EMP DLAs and 2 systems reported in the literature, we conduct a comprehensive analysis of eight of the most metal-poor gas clouds currently known. We focus on high-precision abundance measurements using the elements: C, N, O, Al, Si, and Fe.}
{Our findings indicate increasing evidence of elevated [O/Fe] abundances when [Fe/H]~$<-3$. EMP DLAs are well-modelled with a mean value of [O/Fe]$_{\rm cen} = +0.50 \pm 0.04$ and an intrinsic scatter of $\sigma_{\mathrm{int, [O/Fe]}} = 0.13_{-0.04}^{+0.06}$. While VMP DLAs are well-modelled with [O/Fe]$_{\rm cen} = +0.40 \pm 0.02$ and $\sigma_{\mathrm{int, [O/Fe]}} = 0.06 \pm 0.02$. We further find tentative evidence of a redshift evolution of [C/O] across these most metal-poor DLAs with lower redshift systems showing elevated [C/O] ratios. Using the measured abundances, combined with a stochastic chemical enrichment model, we investigate the properties of the stellar population responsible for enriching EMP gas at cosmic noon. We find that the chemistry of these systems is best explained via the enrichment of just two massive progenitors, $N_{\star} = 2\pm 1$, that ended their lives as core collapse SNe with a typical explosion energy $E_{\rm exp} = (1.6 \pm 0.6)\times10^{51}$~erg. These progenitors formed obeying a Salpeter-like power-law IMF, where all stars of mass greater than $M_{\rm max}=32_{-4}^{+10}~{\rm M_{\odot}}$ collapse directly to black holes and do not contribute to the metal enrichment.}
{}

   \keywords{Damped Lyman alpha systems --
                Intergalactic medium --
                Population III stars --
                Population II stars --
                Chemical abundances
               }

   \maketitle
%

\section{Introduction} \label{ofe:sec:intro}
Tracing chemical evolution from the Cosmic Dawn ($15<z<30$) to the present day unequivocally begins with the first generation of Population III (Pop. III) stars \citep{BrommYoshida2011}. Understanding the properties of these stars is essential to unveiling the first instances of chemical enrichment and the origin of chemical elements in the Universe \citep{Kobayashi2020, Klessen2023}.

Within the current cosmological framework, the primordial Universe was primarily composed of hydrogen, helium, and trace amounts of lithium. The creation of the first metals began with the birth of the first stars; the cores of these Pop. III stars were the dominant source of heavy elements at this epoch.

Current state-of-the-art simulations of Pop. III star formation suggest that these stars may have had a relatively high characteristic mass ($M>10$~M$_{\odot}$) and were therefore short-lived \citep{BrommYoshida2011}. Thus, the probability of directly detecting Pop. III stars in the local Universe is low. However, when the first stars ended their lives, they enriched the surrounding environment with the first instances of metals. The resulting abundances of these metals depend directly on the properties of the first stars (e.g. their initial mass function; IMF, explosion energy, rotation properties etc.). Thus, by identifying the unique chemical fingerprint of the first stars one may indirectly recover some of their properties. Typically this fingerprint is searched for via the metal-poor stars in the halo of the Milky Way (MW) and its surrounding dwarf galaxies \citep[e.g,][]{Bond1980, Beers1985, Aoki2006, Frebel2005, FrebelNorris2015, Hartwig2018Aug, Hartwig2023, Ishigaki2018, Starkenburg2017, Arentsen2020, Vanni2023}. 
It has also been realised that metal-poor gas reservoirs at high redshifts can be used to search for this fingerprint \citep{Erni2006, Pettini2008, Penprase2010, Cooke2017, Welsh2019, Welsh2023, HazeNunez2021}. 

These gas reservoirs are typically detected as absorption along the line-of-sight towards unrelated background quasars. The highest column density gas reservoirs, whose column density of neutral hydrogen exceeds $\log_{10} (N({\rm H}$\,{\sc i}$)/{\rm cm}^{-2})\ge 20.3$ are known as Damped Ly$\alpha$ systems (DLAs; see \citealt{Wolfe2005} for a review). The significant amount of neutral hydrogen in these gas clouds shields the relatively fewer metals from ionizing photons. As a result, most elements associated with the neutral gas are in a single dominant ionization stage. These dominant ionization stages are known based on the relative excitation energies with respect to that of hydrogen; ionization corrections are therefore negligible, provided that the abundance of the dominant ion stage of each metal can be measured. Thus, these DLAs are arguably some of the best astrophysical environments to precisely determine chemical abundances in the early Universe. Also of interest are the slightly lower column density absorbers known as sub-DLAs. These reservoirs have a neutral H\,{\sc i} content between $ 19.0 <\log_{10} (N({\rm H}$\,{\sc i}$)/{\rm cm}^{-2}) < 20.3$ and may intersect partially ionized gas.

Crucially, gas reservoirs detected in absorption allow us to study low density structures in the early Universe that are otherwise challenging to detect. They are an invaluable tool to study the evolution of metals across cosmic time. At redshift $z\sim3$, approximately 60 per cent of observed metals are found in low-ionisation gas often associated with DLAs. This fraction increases at higher redshifts \citep{Peroux2020}.

Of particular interest is the evolution of some of the most abundant chemical elements (e.g. C, O, N, Ne, and Si) alongside that of iron-peak elements (e.g. Fe, Ni, Zn). Historically, searches for signatures of the first stars have focused on discovering the most iron-poor environments. In these investigations, it is typical to exclusively consider systems that are at least extremely metal-poor (EMP). EMP systems are defined as having an iron abundance that is between 1/10\,000$^{\rm th}$ and 1/1000$^{\rm th}$ of the solar abundance \citep{Beers2005}. This is often expressed as\footnote{The square bracket notation denotes the logarithmic number abundance ratio of elements X and H relative to their solar values X$_{\odot}$ and H$_{\odot}$, i.e. $[{\rm X / H}] =  \log_{10}( N_{{\rm X}}/N_{{\rm H}}) - \log_{10} (N_{{\rm X}}/N_{{\rm H}})_{\odot}$.} $-4<$~[Fe/H]~$<-3$. Systems with an [Fe/H] abundance between $-3<\rm [Fe/H] <-2$ are classified as being very metal-poor (VMP).

The behaviour of [$\alpha$/Fe] as a function of the Fe-metallicity is an informative probe of chemical enrichment\footnote{The $\alpha$ elements are those that can be produced via the $\alpha$-capture process (e.g. O, Ne, Mg, Si, S, Ar, Ca).}. The evolution of this relative abundance ratio can indicate the onset of enrichment from Type Ia supernovae (SNe) \citep{Tinsley1979, Wheeler1989}. This was first detected within the MW whose constituent stars exhibit a plateau in the [$\alpha$/Fe] abundance ratio at low metallicities. This persists until [Fe/H]~$\sim -1$, when the [$\alpha$/Fe] abundance begins to decline \citep{Matteucci1990, Matteucci2003, Edvardsson1993, Ishigaki2012}. Since Type Ia SNe predominantly produce Fe-peak elements, this decline in [$\alpha$/Fe] corresponds to the onset of enrichment from long-lived, low-mass stars that left the main sequence as white dwarfs. Prior to the enrichment from these low mass stars, it is the IMF-weighted abundances of the more massive and, consequently, short-lived stars that are being observed. These massive stars are rich in both $\alpha$- and Fe-peak elements.  Once the supply of massive stars is exhausted, there is no way to replenish the $\alpha$-abundances (without a subsequent burst of star-formation).

The stellar populations of nearby satellite dwarf spheroidal galaxies (dSphs) also exhibit this so-called `$\alpha$-knee'. In these more metal-poor environments, the knee occurs at lower metallicities than in the MW \citep{TolstoyHillTosi2009, Kirby2011}. Sculptor and Fornax show a decline in [$\alpha$/Fe] at [Fe/H]~$=-1.8$ and [Fe/H]~$=-1.9$, respectively \citep{Starkenburg2013, Hendricks2014, Hill2018}. 
This behaviour has also been characterised for metal-poor DLAs at $z\sim3$ \citep{CookePettiniJorgenson2015, Velichko2024}. The metallicity of the `$\alpha$-knee' across the metal-poor DLA population is found to be [Fe/H]~$\sim-2$. While this knee is observed across a sample of objects, rather than an individual galaxy, it is still an informative measure of chemical enrichment.

\indent Oxygen is the most abundant metal in the Universe \citep{Lodders2019, Magg2022}. It is a key $\alpha$-element for investigating chemical evolution. VMP DLAs exhibit a plateau in their [O/Fe] abundance at [O/Fe]~$\simeq+0.4$ \citep{Cooke2011b}. This suggests that the DLAs in this metallicity regime were enriched by a similar population of stars drawn from the same IMF. This is comparable to the IMF-weighted abundance of elements observed across galaxies that show an $\alpha$-knee. Notably, the [Fe/H] position of the knee observed for both Sculptor and Fornax coincides with the region in which the DLA abundances appear to initially plateau ([Fe/H]~$\lesssim-2$). 

At lower metallicities, bordering on the EMP regime, this plateau in [O/Fe] appears to break down. Previous works indicate that this ratio is enhanced relative to the higher metallicity systems. This could be a signature of enrichment by a generation of metal-free stars (e.g. \citealt{HegerWoosley2010}). Indeed, theoretical works that explore the Pop. III signature in the intergalactic medium (IGM) at $z\sim3$ find that the imprint of the first stars may be seen in gas reservoirs with [Fe/H]~$<-3$ that additionally show an [O/Fe] enhancement \citep{Kirihara2020}. In fact, some models indicate that the entire abundance patterns observed across metal-poor DLAs at $z\sim 3$ are consistent with Pop. III enrichment \citep{Maio2015, Webster2015b}. 

\begin{table*}
\centering
\caption{Journal of observations associated with our sample \label{tab:sum} 
}
\begin{tabular}{llllllllll}
\hline
QSO & $m_{\rm r}$ & $z_{\rm em}$ & $z_{\rm abs}$ & Telescope/ & Wavelength & Resolution & Integration & S/N$^{b}$ & Programme \\
 & (mag) &  &  & instrument & range$^{a}$ (\AA) & (km\,s$^{-1}$) & time (s) &  & ID \\
 \hline
J0140$-$0839 & 17.75  & 3.72 & 3.697 & VLT/UVES    & 3050 -- 10430  & 8.3   & 12\,000   & 30    & 080.A-0014(A) \\
             &        &       &       & VLT/UVES    & 3760 -- 10430  & 7.3   & 60\,000   & 51    &  105.20L3001   \\

J0239$-$0649   & 18.66  & 2.49 & 2.416 & VLT/UVES    & 3755 -- 9465  & 7.3   & 18\,000   & 11    & 108.222R.001  \\
J0903$+$2628   & 18.98  & 3.22 & 3.078 & Keck/HIRES  & 3840 -- 6715  & 6.3   & 32\,400   & 20    & N162Hb, C230Hb \\
             &        &       &       & Keck/HIRES  & 3245 -- 7805  & 6.3   & 32\,400   & 36    & R351     \\
J1001+0343 & 19.02  & 3.20 & 3.078 & VLT/UVES & 3282 -- 6655  & 7.3 & 33\,700   &  28   & 083.A-0042(A) \\
  &       &       &        & VLT/UVES    & 3756 -- 10429  & 7.3    & 24\,000   & 15    & 105.20L3.001  \\ 
J1147+5034   & 19.45  & 2.52 & 2.519 & Keck/HIRES    & 4070 -- 8760  & 6.3   & 18\,000   & 22    & R304 \\
J1358+0349   & 18.64  & 2.89 & 2.853 & VLT/UVES    & 3285 -- 9460 & 7.3   & 21\,600   &  30    & 111.24LY.001 \\
             &        &       &       & VLT/UVES     & 3450 -- 6645 & 6.4   & 24\,000   & 30    & 093.A-0016(A)  \\ 
             &        &       &       & Keck/HIRES   & 3480 -- 6345 & 6.2   & 24\,000   & -    & H228Hb  \\ 
J2150+0331   & 19.12  & 2.79 & 2.697 & VLT/UVES    &  3760 -- 9465  & 7.3   & 30\,000   & 11    & 110.23R3.002 \\

J2308+0854   & 19.40  & 2.77 & 2.768 & VLT/UVES    & 3285 -- 9465  & 7.3   & 36\,000   & 19   & 108.222R.002 \\
 \hline
\end{tabular}
\begin{tabular}{l}
     $^{a}$ With some wavelength gaps. \\
     $^{b}$ Near the bluest accessible Fe\,{\sc ii} line --- typically $\lambda1608$ or $\lambda2344$. This is left empty if an Fe\,{\sc ii} feature is not covered. In the case of \\ Fe\,{\sc ii}~$\lambda1608$, this is often the weakest line we cover and so it is a useful indicator of the quality of these data.
\end{tabular}
\end{table*}

\indent Discerning the true behaviour of [O/Fe] in the EMP regime is challenging due to the typical size of the associated uncertainties ($>0.1$ dex) compared to the expected strength of the enhancement (variable depending on the underlying IMF). These challenges can be overcome through deep observational programmes curated for absorbers of interest. Recent work, led by this group, has highlighted the value in such a programme when targeting DLAs between $2< z_{\rm abs} < 4$. In particular, our observations and subsequent analysis of an opportunely placed DLA in the [O/Fe]-[Fe/H] parameter space strengthened the observed signature of an elevated [O/Fe] ratio in the EMP regime compared to the VMP regime with a significance of $\sim2\sigma$ \citep{Welsh2022}. This was further reaffirmed by our latest observations and analysis of the most metal-poor DLA currently known \citep{Cooke2017, Welsh2023}.  

\indent Indeed, this previous work also highlights the value of each newly discovered EMP DLA when investigating early chemical enrichment and searching for the chemical fingerprint of the first stars. Discovering this signature is one of the key science goals of the next generation of telescope facilities \citep{Dodorico2023}.

Notably, metal-poor systems, particularly EMP DLAs, make up a small fraction of the overall number of absorbers discovered at $2<z<4$ \citep[known as cosmic noon;][]{Peroux2020}. When looking at statistical surveys, they are often not detected \citep{Rafelski2012, Lehner2022}. However, they can be found through dedicated surveys, as shown in \citet{Pettini2008} and \citet{Penprase2010}. Indeed, there has been increasing interest in studying metal-poor DLAs with high resolution optical spectrographs for more than two decades \citep[e.g,][]{Molaro2000, Dessauges-Zavadsky2001, Omeara2001, Erni2006, Petitjean2008, Cooke2011b, Cooke2014, Dutta2014, Morrison2016, Berg2024}. Cosmic noon is the only epoch thus far where completely chemically pristine gas has been detected \citep{Fumagalli2012, Robert2019}. It is also the epoch in which both the strongest metal absorption features and the associated Ly$\alpha$ absorption (rest frame UV) fall in the optical wavelength range. With the ever-increasing number of quasars being found at redshift $z>6$, it may also become possible to discover pristine gas closer to the epoch of the Cosmic Dawn by searching for proximate absorption line systems \citep[e.g,][]{Banados2019, Sodini2024}.

Here we present the results of our latest survey of metal-poor gas clouds at cosmic noon and investigate their associated chemical abundance pattern using high resolution echelle spectroscopy. Specifically, we report the discovery of four new metal-poor systems. One of these newly discovered systems is a bonafide EMP DLA with [Fe/H]~$=-3.13 \pm 0.06$. Another system is on the cusp of this regime with [Fe/H]~$=-3.00 \pm 0.12$ and the remaining two systems are found to be VMP. Together with these new discoveries, we summarise our ongoing programme designed to obtain new data of known EMP DLAs. We present the results of this bespoke programme in the form of new high resolution data of two previously known EMP DLAs. Combined with the work presented in \citet{Welsh2022, Welsh2023}, we form a sample of 8 of the most metal-poor high column density absorbers at cosmic noon and use these data to determine the precise chemical abundance patterns in this regime. In particular, we focus on the chemical evolution of [O/Fe].

 This paper is organised as follows. Section \ref{sec:obs} describes our targets and the observations. Section \ref{sec:red} details the data reduction and fitting procedure process. In Section \ref{sec:ana}, we present the details of our sample and our data. We analyse the chemical abundances in Section~\ref{sec:chem} and consider possible impacts of physical processes in Section~\ref{sec:phys}. In Section \ref{sec:disc}, we discuss these results and draw comparisons with other works, before drawing overall conclusions and suggesting future work in Section \ref{sec:conc}.

\section{Observations} 
\label{sec:obs}

The data presented in this paper are either the first high-resolution observations of the DLA or we have obtained additional high-resolution observations designed to pin down the Fe\,{\sc ii} column density with increased precision. A summary of the observations can be found in Table~\ref{tab:sum}. These data were taken with either the Ultraviolet and Visual Echelle Spectrograph (UVES; \citealt{Dekker2000}) at the European Southern Observatory (ESO) Very Large Telescope (VLT) or the High Resolution Echelle Spectrometer (HIRES; \citealt{Vogt1994}) at the Keck observatory. Our latest observations of the individual systems are described in turn. Unless otherwise stated, all observations were carried out with a $0.8''$ slitwidth and $2\times2$ binning during readout. 
\\

\subsection{Known EMP DLA programme}
\subsubsection{J0140$-$0839}
The proximate DLA found towards the QSO SDSS J014049.18$-$083942.50 (hereafter J0140$-$0839), with $m_{\rm r}=17.75$, was first identified using data from the Sloan Digital Sky Survey (SDSS) \citep{Prochaska2008}. High resolution echelle spectroscopic data were then taken with UVES ($R\sim 40\,000$) and presented in \citet{Ellison2010}. Based on these data and the subsequent reanalysis presented in \citet{Cooke2011b}, this DLA was identified as one of the most Fe-poor DLAs known with an estimated Fe abundance of [Fe/H]~$=-3.45\pm 0.24$. In this paper, we present an additional $\sim$16 hours of VLT/UVES data taken during observing periods P105 to P108. Approximately 8 hours of observations were taken with the 437$+$760 setup and a further $\sim$8 hours were taken with the 437$+$860 setup, both executed as $10\times 3000$~s exposures. 

\subsubsection{J1358$+$0349}
The DLA towards the quasar QSO SDSS J135803.97$+$034936.01 (hereafter J1358$+$0349) with $m_{\rm r} = 18.64$ was, at one point, the most metal-poor DLA known. It has been observed extensively to study the primordial abundance of deuterium \citep{Cooke2016}. Consequently, previous observations focused on blue wavelengths. We present new VLT/UVES data that focus on red wavelengths and cover the previously unobserved \fe{2344} and \fe{2382} features. These intrinsically strong absorption lines fall at red wavelengths that are commonly impacted by telluric features. To accommodate for this potential contamination, we observed a telluric standard star (using an identical instrument setup) alongside every exposure on target. This allows for the accurate removal of telluric features from the spectrum extracted from each exposure. The process of telluric removal can be found in Appendix~\ref{appen:skylines} alongside additional tests.  The latest data were taken with the 390$+$760 setup and executed as $8\times 2700$~s exposures. All exposures were followed with a 3~s exposure of a nearby A0V star, HD 126129. For the details of the archival UVES and HIRES data we refer the reader to \citet{Cooke2016}. \\

\subsection{Survey for new systems}
 \noindent In addition to our follow-up survey to study all known EMP DLAs with high-precision, we present the latest results from our ongoing search for new DLAs in the EMP regime. These DLAs were initially identified from SDSS spectroscopic data using the \citet{Parks2018} DLA catalogue. Using this catalogue, we identified a sample of candidate DLAs that appeared to be at least VMP. We sought additional data of these targets through dedicated observations with an intermediate resolution spectrograph; specifically, the recently decommissioned Intermediate-dispersion Spectrograph and Imaging System (ISIS) at the  4.2~m William Herschel Telescope (WHT). The resolution of these follow-up data was sufficient to identify the most promising candidates for further observations with a high resolution echelle spectrograph. In particular, we focus on absorption line systems that would be of interest for studies of near-pristine gas. Thus, we acquired VLT/UVES  or Keck/HIRES data to pin down the detailed chemistry and kinematics of these absorption line systems. The details of these observations are below. In all cases, we  are reporting the identified absorber for the first time. Further, these are the first observations of each system obtained with sufficient resolution to resolve the cloud properties. 
 
 \subsubsection{J0239$-$0649}
Based on observations with VLT/UVES, we have identified the DLA towards the $z_{\rm em}= 2.49$ $m_{r}= 18.66$ QSO SDSS J02396.99$-$064931.15 (hereafter J0239$-$0649) at $z_{\rm abs}=2.416$. We observe the sightline using the 437$+$760 setup for a total of 18\,000~s. These observations were executed as $6\times3000$~s exposures and reach a final combined S/N near the bluest accessible Fe\,{\sc ii} feature of S/N~$=11$.
 
 \subsubsection{J1147$+$5034}
 Through observations with Keck/HIRES, we have identified the proximate DLA towards the $z_{\rm em}=2.52$ $m_{\rm r} = 19.45$ QSO SDSS J114745.03+503446.8 (hereafter J1147+5034) with $z_{\rm abs} = 2.519$. This sightline was observed using the red cross disperser, a cross disperser angle of 0.43 degrees, and an echelle angle of 0.58 degrees combined with a slit length of $7.0''$ and a slit width of $0.861''$ (decker C1). The measured resolution of these data is R~$\simeq50,000$ ($\equiv 6$~\kms\ full width at half maximum; FWHM). We executed our observations as 5$\times$3600~s exposures. With these data, we reach a final combined S/N near the accessible Fe\,{\sc ii} feature of S/N~$=22$.
 
 \subsubsection{J2150$+$0331}
 The third newly observed DLA is found towards the $z_{\rm em}=2.79$ $m_{\rm r} = 19.12$ QSO SDSS J215037.68$+$033131.39 (hereafter J2150$+$0331) at $z_{\rm abs} = 2.697$. This sightline was observed with VLT/UVES in P110 using the 437+760 setup for a total of 30\,000~s executed as 10$\times$3000~s exposures. With these data, we reach a final combined S/N ratio near the accessible Fe\,{\sc ii} feature of S/N~$=11$.
 
 \subsubsection{J2308$+$0854}
 The final absorption line system presented in this paper is found towards the $z_{\rm em}=2.77$ $m_{\rm r} = 19.40$ QSO SDSS J230844.19$+$085442.28 (hereafter J2308$+$0854) at $z_{\rm abs}=2.768$. This sightline was observed with VLT/UVES throughout P108 using the 390+564 setup as well as the 437+760 setup. In total, we obtained 12$\times 3000$~s exposures; 3 of these were executed with the 390+564 setup with the primary aim of observing the absorption due to neutral hydrogen. The remaining 9 exposures were executed using the 437+760 setup to cover the majority of the metal lines of interest.  With these data, we reach a final combined S/N ratio near the accessible Fe\,{\sc ii} feature of S/N~$=19$.

\section{Data reduction and analysis}
\label{sec:red}
\subsection{Data reduction}

The HIRES data were reduced using two different software packages. For data taken prior to 2022, the Hires REDUX\footnote{Hires REDUX is available from: \\ \url{https://www.ucolick.org/~xavier/HIRedux/}} reduction pipeline was used \citep{Bernstein2015}. For data taken after this epoch, the data were reduced with the {\ttfamily{PypeIt}}\footnote{Pypeit is available from: \\ \url{https://pypeit.readthedocs.io/en/release/index.html}} reduction pipeline \citep{Prochaska2020}. The ESO data were all reduced with the EsoRex reduction pipeline. 
Each pipeline includes the standard reduction steps of subtracting the detector bias, locating and tracing the echelle orders, flat-fielding, sky subtraction, optimally extracting the 1D spectrum, and performing a wavelength calibration. The data were converted to a vacuum and heliocentric reference frame.

Finally, we combined the individual exposures of each DLA using {\sc uves\textunderscore popler}\footnote{{\sc uves\textunderscore popler} is available from: \\\url{https://github.com/MTMurphy77/UVES\textunderscore popler}.}$^{,}$\footnote{The exception to this is the J1147+5034 Keck/HIRES data. These data were combined with the {\ttfamily{PypeIt}} coadding tool.} \citep{Murphy2019}.
This corrects for the blaze profile, and allowed us to manually mask cosmic rays and minor defects from the combined spectrum. 
When combining these data we adopt a pixel sampling of 2.5~km\,s$^{-1}$. In some cases, there are differences in resolution across the available data associated with a given DLA. This is a result of observations being taken with different slitwidths. When this is the case, we combine and analyse the data separately. We further note that some of these data were taken $>10$~years apart. To avoid issues associated with combining data of quasars with significant continuum variability, we similarly treat data taken across substantially different periods (i.e., in excess of 2 years) separately.

\subsection{Analysis approach}
\label{sec:anaapp}

\begin{table*}
	\centering
	\caption{Ion column densities of the DLA at $z_{\rm abs}= 3.696621 \pm 0.000002$ towards the quasar J0140$-$0839. The quoted column density errors are the $1\sigma$ confidence limits. The total column density of hydrogen can be found in Table~\ref{tab:chem}.
 }
	\label{tab:j0140}
	\tabcolsep=0.10cm
\begin{tabular}{cccccccc}
\hline 

& & & & $\log_{10} N$(X)/cm$^{-2}$ & & &  \\
Comp. & $z_{\rm abs}$ & $b_{\rm turb}$ [\kms] & C\,{\sc ii} & O\,{\sc i} & Al\,{\sc ii} & Si\,{\sc ii} & Fe\,{\sc ii} \\
\hline
1 & 3.696253 $\pm$ 0.000005  & 0.001 (fixed)$^{a}$ & --- & 13.16 $\pm$ 0.04 & --- & 11.68 $\pm$ 0.07 & ---  \\ 
2 & 3.69642 $\pm$ 0.00002  & 10.7 $\pm$ 1.0 & 13.35 $\pm$ 0.05 & 13.48 $\pm$ 0.07 & 11.23 $\pm$ 0.06 & 12.50 $\pm$ 0.05 & ---  \\ 
3 & 3.696621 $\pm$ 0.000002  & 5.45 $\pm$ 0.23 & 13.62 $\pm$ 0.03 & 14.31 $\pm$ 0.01 & 11.33 $\pm$ 0.04 & 12.90 $\pm$ 0.02 & 12.64 $\pm$ 0.05  \\ 
4 & 3.697022 $\pm$ 0.00002  & 15.7 $\pm$ 2.4 & 13.32 $\pm$ 0.07 & 13.97 $\pm$ 0.06 & 11.21 $\pm$ 0.09 & 12.64 $\pm$ 0.06 & ---  \\ 
5 & 3.697283 $\pm$ 0.000001  & 5.65 $\pm$ 0.29 & 13.58 $\pm$ 0.03 & 14.25 $\pm$ 0.03 & 11.42 $\pm$ 0.04 & 12.88 $\pm$ 0.03 & 12.56 $\pm$ 0.06  \\ 
6 & 3.697629 $\pm$ 0.000003  & 4.30 $\pm$ 0.41 & 12.84 $\pm$ 0.04 & 13.17 $\pm$ 0.03 & 10.83 $\pm$ 0.08 & 12.17 $\pm$ 0.02 & ---  \\ 
7 & 3.69797 $\pm$ 0.00002  & 15.1 $\pm$ 2.2 & 12.92 $\pm$ 0.05 & --- & --- & --- &    --- \\ 
\end{tabular}
\begin{tabular}{l}
     $^{a}$ This represents the line broadening due to the turbulence of this component only. The total Doppler broadening is a combination of \\ the  thermal and turbulent broadening ($b_{\rm tot}^{2} =  b_{\rm th}^{2} + b_{\rm turb}^{2}$). In this case $T=1.0\times10^{4}$~K. \\

\end{tabular}
\end{table*}

\indent Using the Absorption LIne Software ({\sc alis}) package\footnote{{\sc alis} is available from:\\ \url{https://github.com/rcooke-ast/ALIS}.}--- which uses a $\chi$-squared minimisation procedure to find the model parameters that best describe the input data --- we simultaneously analyse the full complement of high S/N and high spectral resolution data currently available for each DLA. We model the absorption lines with Voigt profiles, which consist of three free parameters: a column density, a redshift, and a line broadening parameter. We assume that all lines of comparable ionization level have the same redshift, and any absorption lines that are produced by the same ion all have the same column density and broadening prescription. The total broadening of the lines includes a contribution from both turbulent and thermal broadening. The turbulent broadening is assumed to be the same for all absorption features, while the thermal broadening depends inversely on the square root of the ion mass; thus, heavy elements (e.g. Fe) will exhibit absorption profiles that are intrinsically narrower than the profiles of lighter elements, (e.g. C). There is an additional contribution to the apparent line broadening due to the instrument; we approximate the instrument broadening function using a Gaussian with a width that depends on the instrument setup. For the HIRES and UVES data, the nominal instrument resolutions are $v_{\rm FWHM}= 6.28$~km~s$^{-1}$ (HIRES C1), $v_{\rm FWHM}= 8.33$~km~s$^{-1}$ (UVES slitwidth=1"), and $v_{\rm FWHM} =7.3$~km~s$^{-1}$ (UVES slitwidth = 0.8").
Finally, we note that we simultaneously fit the absorption and quasar continuum of the data. We model the continuum around every absorption line as a low-order Legendre polynomial (typically of order 3). We assume that the zero-levels of the sky-subtracted UVES and HIRES data do not depart from zero\footnote{We visually inspected the troughs of saturated absorption features to confirm this is the case.}. 
This approach ensures that the final ion column densities and their associated errors capture the additional correlated uncertainties associated with the localised continuum model. 

In the following section, we discuss our sample in detail and present the results of this profile fitting for each DLA. We first attempt to use the full complement of available data to determine the gas temperature and turbulent broadening alongside the ion column densities. In the case of clouds with multiple components, it is not possible to assess the turbulent and thermal contribution to the overall Doppler broadening. Therefore, in these instances, we adopt cloud models that assume a temperature typically observed in H~{\sc i} dominated gas \citep{Cooke2014, Noterdaeme2021}. The range of temperatures we consider is $T=(0.8 -1.2)\times 10^{4}$~K. If the resulting column densities are consistent within this temperature range, we report column densities and turbulent broadening under the assumption of $T=1\times10^{4}$~K. We also adopt this approach if the turbulent motions of the gas are significantly dominant over the potential thermal contribution and if the level of saturation makes our fiducial approach untenable. This approach does not impact the derived column densities since the absorption features that are covered for the ions of interest are typically weak and unsaturated.

We present the column densities of all detected components in Tables~\ref{tab:j0140} - \ref{tab:j2308} for each gas cloud. Here, and subsequently, the errors are given by the square root of the diagonal term of the covariance matrix calculated by {\sc alis} at the end of the fitting procedure. To accommodate the presence of potentially ionised gas, we pay particular attention to the components that are traced by neutral oxygen since these should be tracing the predominantly neutral gas. For the case of complex systems, the components traced by both O and Fe are highlighted appropriately in the relevant figures. We adopt a strategy of marking the neutral components with red ticks. These are consistently also the strongest components. Indeed, when considering the chemistry of these clouds, we also restrict our analysis to the components traced by neutral gas.

\section{Sample} 
\label{sec:ana}
In this section we present the details of our sample and the results of our line fitting procedure. The relative abundances of the DLAs in our sample will be discussed in Section~\ref{sec:chem} and are summarised in Table~\ref{tab:chem}. Readers interested in the relative abundances of the sample, and not the details on individual systems, can refer to Section \ref{sec:chem}. 
\subsection{Known EMP DLAs}
\label{sec:known}
Half of our sample is comprised of the four previously known EMP DLAs that we will now discuss in turn.

\subsubsection{J0140$-$0839}

\begin{figure*}
    \centering
    \includegraphics[width=0.9\textwidth]{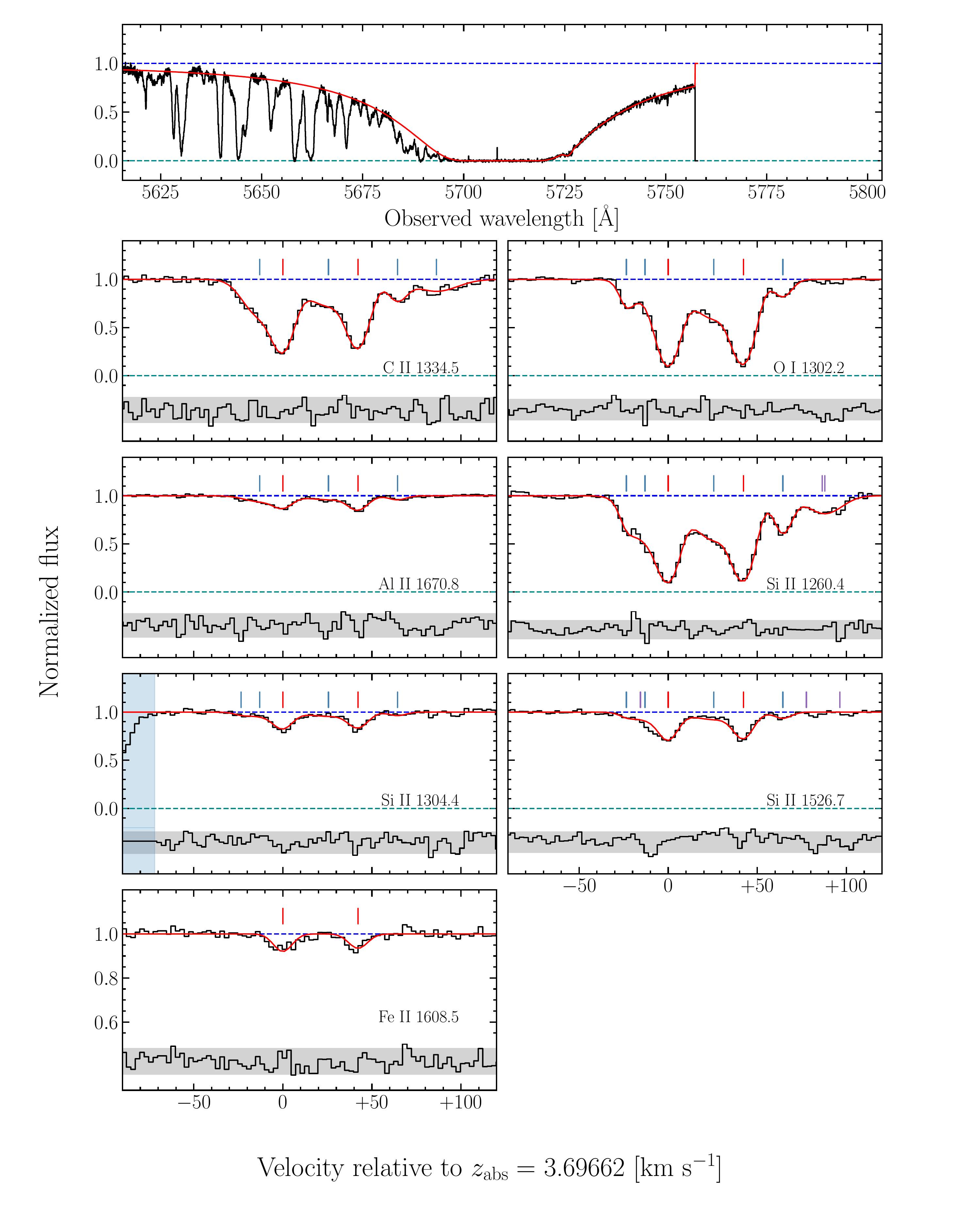}
    \caption{Continuum normalized UVES data (black histograms) of the absorption features associated with the DLA at $z_{\rm abs} = 3.69662$ towards J0140$-$0819. The best-fitting model is shown with the red curves. The blue dashed line indicates the position of the continuum while the green dashed line indicates the zero-level. The ticks above the absorption features indicate the centre of the Voigt line profiles. The features of the strongest components are distinguished via red ticks while those of weaker components are shown in blue. These strongest components are also often associated with the predominantly neutral gas. The purple ticks indicate absorption features unrelated to the DLA. Below the zero-level, we show the residuals of this fit (black histogram) where the grey shaded band encompasses the $2\sigma$ deviates between the model and the data. The vertical blue shaded bands indicate the regions of the spectrum not included in the fit. Note that the full complement of available data were used in calculating the best-fit model, while we only show a selection of the data in this figure to showcase the model and the data. We also note that the y-axis scale is different in some panels to highlight the very weak absorption lines.
    \label{fig:j0140}}
\end{figure*}

\begin{figure*}
    \centering
    \includegraphics[width=0.9\textwidth]{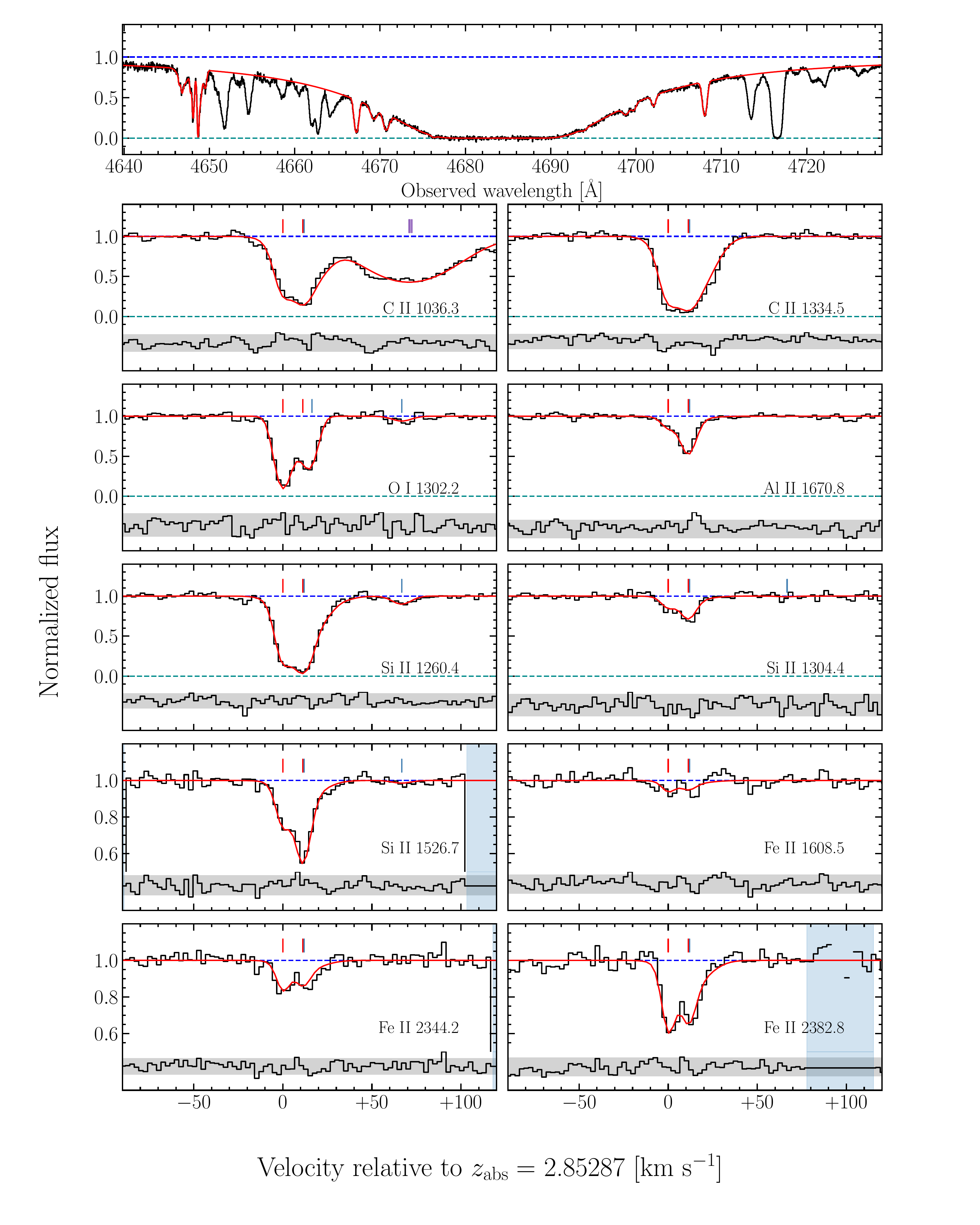}
    \caption{Same as Figure~\ref{fig:j0140} for the DLA at $z_{\rm abs} = 2.85287$ towards J1358$+$0349. Note the different y-scale of the bottom four panels. When fitting hydrogen (top panel) we often fit a selection of unrelated narrow absorption features simultaneously. We only fit those necessary to best capture the shape of the continuum alongside the wings of the damped Ly$\alpha$ profile.}
    \label{fig:j1358}
\end{figure*}

{\footnotesize 
\begin{table*}
	\centering
   \caption{Same as Table~\ref{tab:j0140} for the DLA at $z_{\rm abs}=2.852871\pm 0.000002$ towards the quasar J1358$-$0349. } 
	\label{tab:j1358}
	\tabcolsep=0.10cm
\begin{tabular}{cccccccc}
\hline 
& & & & $\log_{10} N$(X)/cm$^{-2}$ &  & &  \\ 
Comp. & $z_{\rm abs}$ & $b_{\rm turb}$ [\kms] &C\,{\sc ii} & O\,{\sc i} & Al\,{\sc ii} & Si\,{\sc ii} & Fe\,{\sc ii} \\
\hline
1 & 2.852871 $\pm$ 0.000002  & 2.63 $\pm$ 0.33 & 13.48 $\pm$ 0.04 & 14.29 $\pm$ 0.04 & 11.25 $\pm$ 0.06 & 12.64 $\pm$ 0.02 & 12.31 $\pm$ 0.04  \\ 
2 & 2.853015 $\pm$ 0.000002  & 3.49 $\pm$ 0.25 & 13.53 $\pm$ 0.06 & 13.59 $\pm$ 0.10 & 11.94 $\pm$ 0.03 & 13.01 $\pm$ 0.02 & 12.15 $\pm$ 0.09  \\ 
3 & 2.853024 $\pm$ 0.000006  & 13.5 $\pm$ 0.4 & 13.89 $\pm$ 0.03 & --- & 11.10 $\pm$ 0.26 & 12.65 $\pm$ 0.04 & 12.27 $\pm$ 0.11  \\ 
4 & 2.853081 $\pm$ 0.000009  & 1.79 $\pm$ 1.52 & --- & 13.64 $\pm$ 0.07 & --- & --- & ---  \\ 
5 & 2.853728 $\pm$ 0.000010  & 5.55 $\pm$ 1.47 & --- & 12.70 $\pm$ 0.12 & --- & 11.55 $\pm$ 0.07 & ---  \\ 
\hline
\end{tabular}
\end{table*}}

Our latest analysis of J0140$-$0839 indicates that this absorption line system is best modelled by 7 absorption features spread over $\sim$~200 \kms. The column densities of these components are presented in Table~\ref{tab:j0140} alongside their individual redshifts and turbulent broadening. The reduced data and the associated model fit to these data are shown in Figure~\ref{fig:j0140}. The data in this figure are centred on the strongest component at $z_{\rm abs}= 3.696621  \pm 0.000002$. We have refit the Ly$\alpha$ profile associated with this DLA. 
Given this latest model of the continuum, we find log$_{10}$ $N$(H\,{\sc i})/cm$^{-2}~= 20.81 \pm 0.05$. 

While this is a 7 component system, the cloud is dominated by the two strongest components. These strong components are indicated with red ticks in Figure~\ref{fig:j0140}. Indeed, the accessible Fe\,{\sc ii} features are the weak $\lambda$1260 and $\lambda$1608 features and they are both only detected for the two strongest components. Notably, this DLA remains one of the most Fe-poor systems currently known with [Fe/H]~$=-3.37\pm0.06$. This abundance ratio is now known with a three times greater precision compared to previous determinations. This was predominantly facilitated by the higher S/N data available near the \fe{1608} feature. The associated [O/Fe] abundance is now known with four times greater precision. We find [O/Fe]~$=+0.45 \pm 0.04$ while it was previously determined to be [O/Fe]~$=+0.70 \pm 0.19$.

\subsubsection{J0903$+$2628}
The DLA found towards the quasar SDSS J090333.55$+$262836.3 (hereafter J0903$+$2628) is the \emph{most} metal-poor DLA currently known \citep{Cooke2017, Welsh2023}. We recently acquired an additional 10 hours of echelle spectroscopic data on the associated quasar using Keck I/HIRESr. We refer the reader to these papers for details and we present the relevant abundances in Table~\ref{tab:chem}.

\subsubsection{J1001$+$0343}
 The DLA towards SDSS J100151.45+034301.4 (hereafter J1001$+$0343) is a previously known EMP DLA. The VLT/UVES data that facilitated the latest investigation of [O/Fe] are presented in \citet{Welsh2022}. In summary, the latest data reduced the error associated with the [O/Fe] determination by a factor of three. We refer the reader to this work for further details. This earlier work highlights that the behaviour of [O/Fe] in the EMP regime is more easily studied with high precision abundance determinations. Again, the relevant abundances are collected in Table~\ref{tab:chem}. For the first time, we report the [Al/H] abundance of this system.

\subsubsection{J1358$+$0349}
J1358$+$0349 is another system that is best-modelled using multiple components. The reduced data and the associated model fit to these data are shown in Figure~\ref{fig:j1358} while the decomposed column densities of each component are presented in Table~\ref{tab:j1358}. The strongest O\,{\sc i} component is located at  $z_{\rm abs} = 2.852871 \pm 0.000002$. As has been found in previous works, this DLA has a deceptively complex kinematic structure. It cannot be well-modelled assuming that the components share the same relative abundances of metals.

Two components are traced by both O and Fe. These are marked with red ticks in Figure~\ref{fig:j1358}. The central C, Al, Si, and Fe features follow the same 3 components. The main O\,{\sc i} structure is also traced by three components. We note that the first two O\,{\sc i} components are consistent with those found for the singly ionised features. However, the third component shows a small offset to that of the third component traced by C, Al, Si, and Fe. There is a further (fourth) O\,{\sc i} component that is offset from the central cloud at $v\sim+80$~\kms; this is also traced by Si\,{\sc ii}.

Despite this complexity, we can report the total column density of iron for this system using 3 Fe\,{\sc ii} features (specifically $\lambda$1608, $\lambda$2344, and $\lambda$2382) that have different oscillator strengths. With these data, we find [Fe/H]~$=-3.45\pm0.04$. This is based on the  Fe\,{\sc ii} column densities of the two components also traced by O\,{\sc i}. The associated [O/Fe] abundance is [O/Fe]~$=+0.60\pm 0.06$. This is consistent with previous observations and the associated errors have been reduced by a factor of 3. We cannot improve on the neutral hydrogen column density log$_{10}$ $N$(H\,{\sc i})/cm$^{-2}~= 20.524 \pm 0.006$ reported in \citet{Cooke2016}.
{\footnotesize \begin{table}
	\centering
	\caption{Ion column densities of the DLA at $z_{\rm abs}=2.41623$ towards the quasar J0239$-$0649. The quoted column density errors are the $1\sigma$ confidence limits.}
	\label{tab:j0239}
	\tabcolsep=0.10cm
\begin{tabular}{lcccc}
\hline
Ion        & Transitions & $\log_{10} N$(X)/cm$^{-2}$ \\ 
  & used [\AA] &  &  &  \\ \hline \hline
H\,{\sc i}   & 1215             & 20.10 $\pm$ 0.05      \\
C\,{\sc ii}  & 1334       & 14.13 $\pm$ 0.08  \\
N\,{\sc i}  & 1200      & 12.60 $\pm$ 0.16  \\
O\,{\sc i}   & 1302       & 14.72 $\pm$ 0.09    \\
Al\,{\sc ii}   & 1670       & 11.80 $\pm$ 0.12    \\
Si\,{\sc ii} & 1260, 1304,  & 13.40 $\pm$ 0.05   \\
Fe\,{\sc ii} &   2344, 2382, 2586, 2600     & 12.92 $\pm$ 0.03    \\ 

\hline
\end{tabular}
\end{table}}

\subsection{New high column density absorbers}
\label{sec:new}
We now present the data associated with the newly discovered high column density absorbers. As previously discussed, these absorbers were first identified as potentially metal-poor systems through SDSS data and the \citet{Parks2018} DLA catalogue. They remained promisingly metal-poor when observed with an intermediate resolution spectrograph. These data were collected with WHT/ISIS \citep[recently decommisioned in favour of the WHT Enhanced Area Velocity Explorer (WEAVE) survey instrument;][]{Dalton2012}. The targets that appeared to be the most metal-poor based on the ISIS data were then followed up with a high resolution echelle spectrograph capable of resolving the detailed structure of the absorption lines, thereby allowing us to pin down the underlying cloud model and component column densities with confidence.

\begin{figure*}
    \centering
    \includegraphics[width=0.9\textwidth]{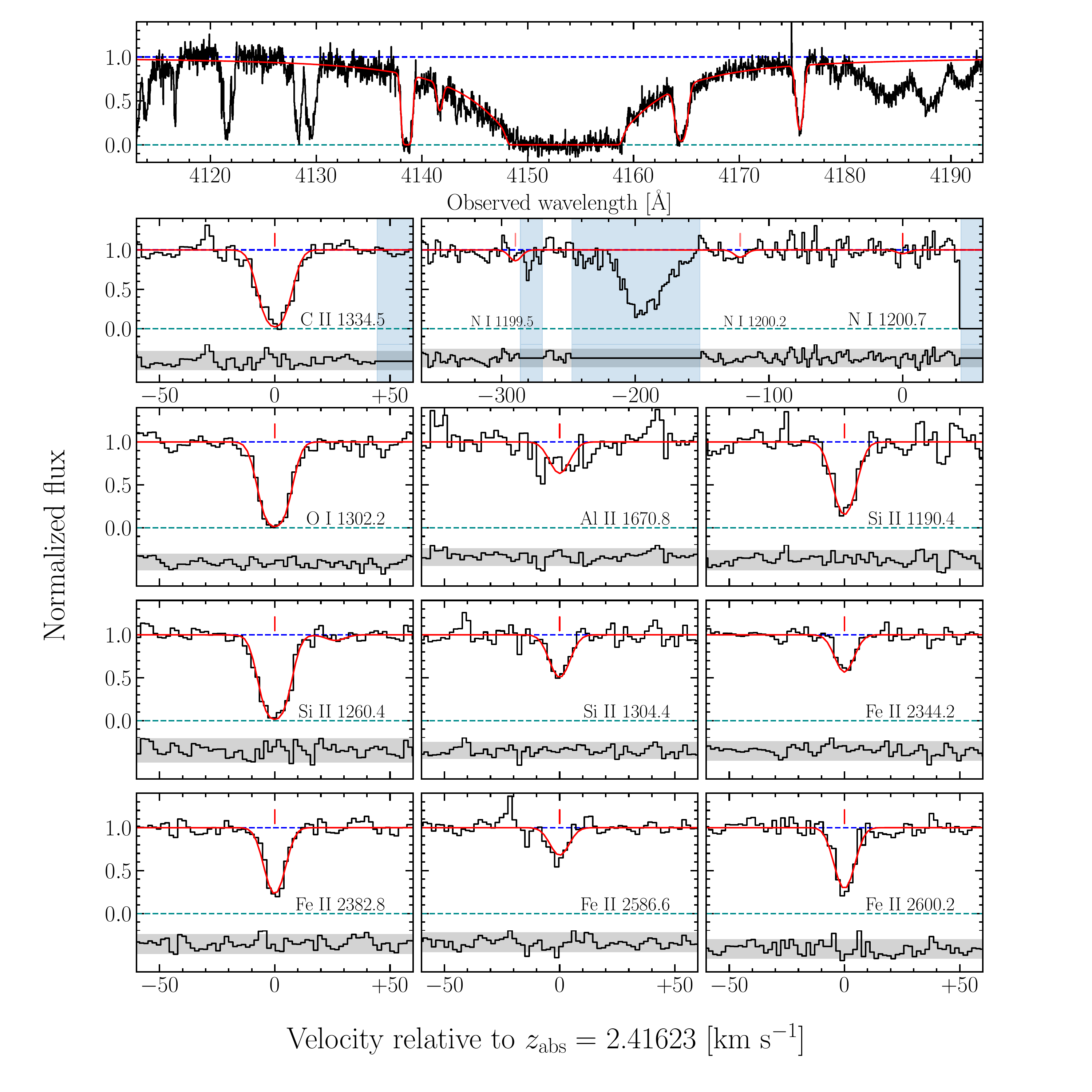}
    \caption{Same as Figure~\ref{fig:j0140} for the DLA at $z_{\rm abs}=2.41623$ towards the quasar J0239$-$0649. Note the x-axis of the panel showing N\,{\sc i} is different to that showing the other ions. This is to showcase the three features of the N\,{\sc i}~$\lambda1200$ triplet. }
    \label{fig:j0239}
\end{figure*}

\subsubsection{J0239$-$0649}
\noindent The absorber towards J0239$-$0649 is best modelled with one gaseous component at $z_{\rm abs}= 2.416233 \pm 0.000001$ for all detected species. The detected features of the light atomic elements C and O are unfortunately saturated. Our analysis is restricted to those with the largest oscillator strengths. The weaker features are covered by our observations but are heavily blended with unrelated absorption.  
Given this saturation, there is not enough information in these data to constrain the turbulent and thermal broadening of the gas simultaneously. Instead, we explore the cloud models that assume a temperature typically observed in H~{\sc i} dominated gas (as discussed in Section~\ref{sec:anaapp}). Under the assumption $T_{\rm}=1 \times 10^{4}~{\rm K}$, we find that the turbulent velocity contribution is $b=3.6\pm 0.2$~km\,s$^{-1}$. The data, along with the best-fitting model are presented in Figure \ref{fig:j0239}, while the corresponding column densities are listed in Table \ref{tab:j0239}. Despite the saturation of the C\,{\sc ii} and O\,{\sc i} lines, one can still obtain reliable abundance measures by using the Si\,{\sc ii} and Fe\,{\sc ii} features, that cover a range of $f\lambda$ values, to pin down the cloud model. There is an additional uncertainty associated with fitting modestly saturated lines since the observed line profiles are less visibly impacted by changes in column density. This can be minimised by fitting all of the metal lines simultaneously. 
\begin{figure*}
    \centering
    \includegraphics[width=0.9\textwidth]{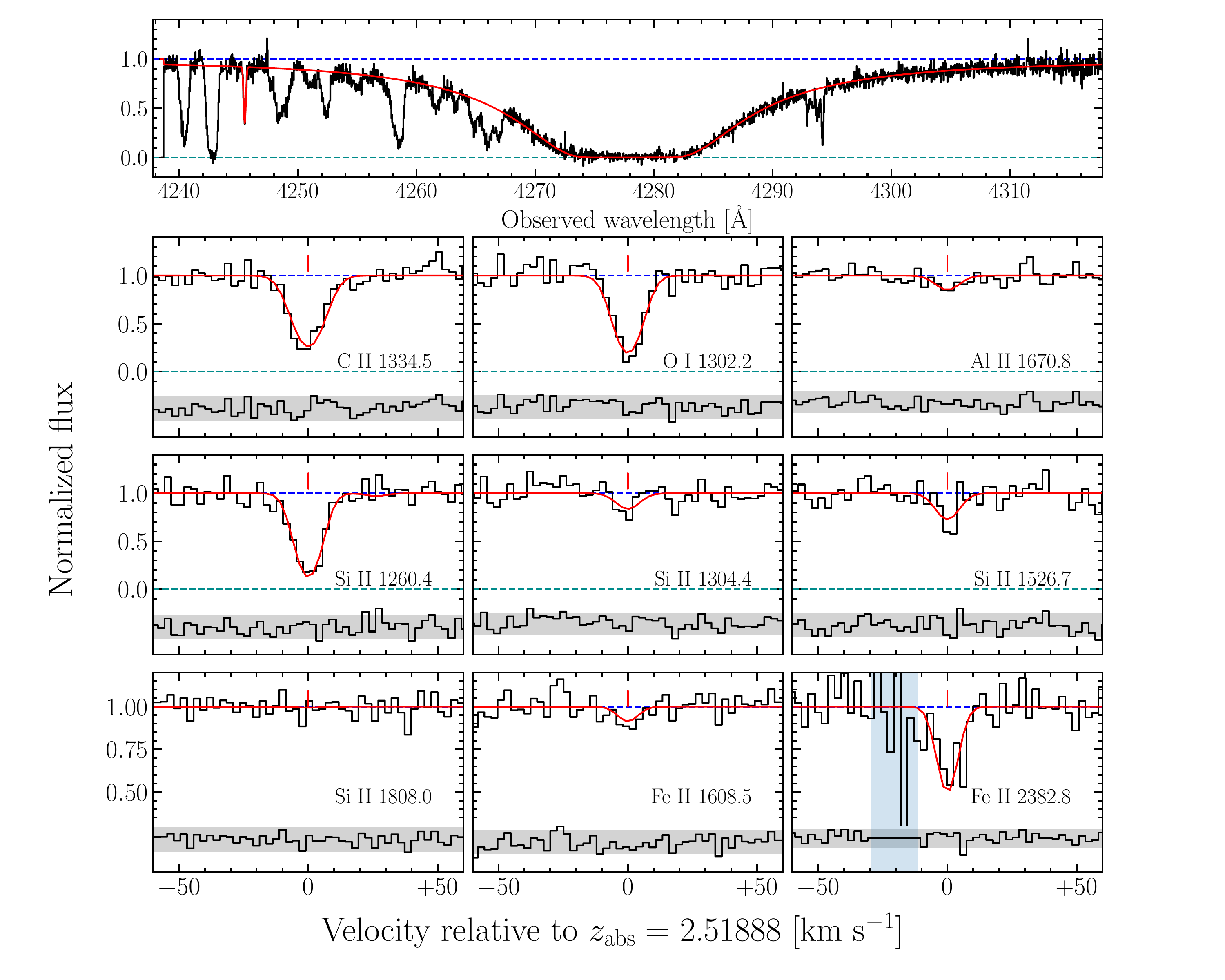}
    \caption{Same as Figure~\ref{fig:j0239} for the DLA at $z_{\rm abs}=2.51888$ towards J1147$+$5034. Note the different y-axis scale of the fourth row. Note also the Si\,{\sc iii}~$\lambda1206$ absorption feature in the top panel at $\sim4246$~\AA. Also note that the Si\,{\sc ii}~$\lambda$1808 feature is not detected. Combined with the detection of stronger Si\,{\sc ii} lines, the coverage of this feature is useful to determine the total Si\,{\sc ii} column density. }
    \label{fig:j1147}
\end{figure*}

We measure [Fe/H]~$=-2.64 \pm 0.06$ placing it firmly in the VMP regime. We find that the total column density of neutral hydrogen is log$_{10}$ $N$(H\,{\sc i})/cm$^{-2}~= 20.10 \pm 0.05$. We do not detect appreciable absorption from any Fe-peak elements beyond Fe (such as Cr, Ni, or Zn). However, we tentatively detect N\,{\sc i} via the $\lambda$1200 triplet with a $\sim2\sigma$ confidence. 
This suggests [N/O]~$\simeq -1.2$ while [N/H]~$= -3.33 \pm 0.16 $. This DLA is therefore found to be inline with the [N/$\alpha$] plateau observed at [N/$\alpha$]~$=-1.45 \pm 0.05$ for DLAs with low [N/H] abundances \citep{Centurion2003, Zafar2014}. It is also in support of works that find that the distribution of [N/O] measured across high redshift absorbers shows a high level of dispersion \citep{Pettini2008, Petitjean2008, Vangioni2018}. Though, we note that the quality of these data are only sufficient to report a marginal N detection. Thus, we defer to future observations for further analysis.

\subsubsection{J1147$+$5034}
The DLA towards J1147$+$5034 is best modelled by one gaseous component at $z_{\rm abs} = 2.518882 \pm 0.000001$. This gas cloud has a temperature of $T = (1.7 \pm 0.8)\times10^{4}$~K and a turbulent doppler parameter of $b = 3.65 \pm 0.91$~\kms. The data, along with the best-fitting model are presented in Figure \ref{fig:j1147}, while the corresponding column densities are listed in Table \ref{tab:j1147}. We find that log$_{10}$ $N$(H\,{\sc i})/cm$^{-2}~= 20.23 \pm 0.01$ and [Fe/H]~$=-3.13\pm0.06$. This is the most metal-poor newly discovered DLA from this survey. It is additionally the lowest redshift EMP DLA discovered to date with an absorption redshift $z_{\rm abs}=2.519$.

As can be seen in the top panel of Figure~\ref{fig:j1147}, these data provide a clean detection of the Si\,{\sc iii} $\lambda$1206 feature. Given that this system has a H\,{\sc i} column density that lies near the border of being classified as a DLA, we can use the successive ion ratio Si\,{\sc iii}/Si\,{\sc ii} to investigate the potential need for ionization corrections (as done in Section~\ref{sec:ioncorr}). We also detect C\,{\sc iv} that is coincident in velocity space with the low ionisation absorption (as reported in Table~\ref{tab:j1147}).

{\footnotesize \begin{table}
	\centering
	\caption{Same as Table \ref{tab:j0239} for the DLA at $z_{\rm abs}=2.51888$ towards the quasar J1147$+$5034. }
	\label{tab:j1147}
	\tabcolsep=0.10cm
\begin{tabular}{lccc}
\hline
Ion        & Transitions & $\log_{10} N$(X)/cm$^{-2}$  \\ 
  & used [\AA] &  &  \\ \hline \hline
H\,{\sc i}   & 1215             & 20.23 $\pm$ 0.01 \\
C\,{\sc ii}  & 1334       & 13.64 $\pm$ 0.03   \\
C\,{\sc iv}  & 1548,1550       & 12.87 $\pm$ 0.03   \\
O\,{\sc i}   & 1302       & 14.12 $\pm$ 0.03    \\
Al\,{\sc ii}   & 1670       & 11.37 $\pm$ 0.10    \\
Si\,{\sc ii} & 1260, 1304, 1526, 1808  & 12.83 $\pm$ 0.03    \\
Si\,{\sc iii} & 1206  & 12.45 $\pm$ 0.02    \\
Fe\,{\sc ii} &   1608, 2382    & 12.56 $\pm$ 0.06   \\ 
\hline
\end{tabular}
\end{table}}

{\footnotesize \begin{table}
	\centering
	\caption{Same as Table \ref{tab:j0239} for the DLA at $z_{\rm abs}=2.69712$ towards the quasar J2150$+$0331.}
	\label{tab:j2150}
	\tabcolsep=0.10cm
\begin{tabular}{lccc}
\hline
Ion        & Transitions & $\log_{10} N$(X)/cm$^{-2}$  \\ 
  & used [\AA] &  &  \\ \hline \hline
H\,{\sc i}   & 1215             & 19.68 $\pm$ 0.05 \\
C\,{\sc ii}  & 1036, 1334       & 13.75 $\pm$ 0.06  \\
C\,{\sc iv}  & 1548, 1550       & 13.50 $\pm$ 0.10   \\
O\,{\sc i}   & 1302       & 14.32 $\pm$ 0.10    \\
Al\,{\sc ii}   & 1670       & 11.37 $\pm$ 0.10   \\
Si\,{\sc ii} & 1260, 1304, 1808  & 12.85 $\pm$ 0.14    \\
Fe\,{\sc ii} &   1260, 1608, 2382    & 12.56 $\pm$ 0.06    \\ 
\hline
\end{tabular}
\end{table}}

\subsubsection{J2150$+$0331}
\begin{figure*}
    \centering
    \includegraphics[width=0.9\textwidth, trim={0 0 0 0}, clip]{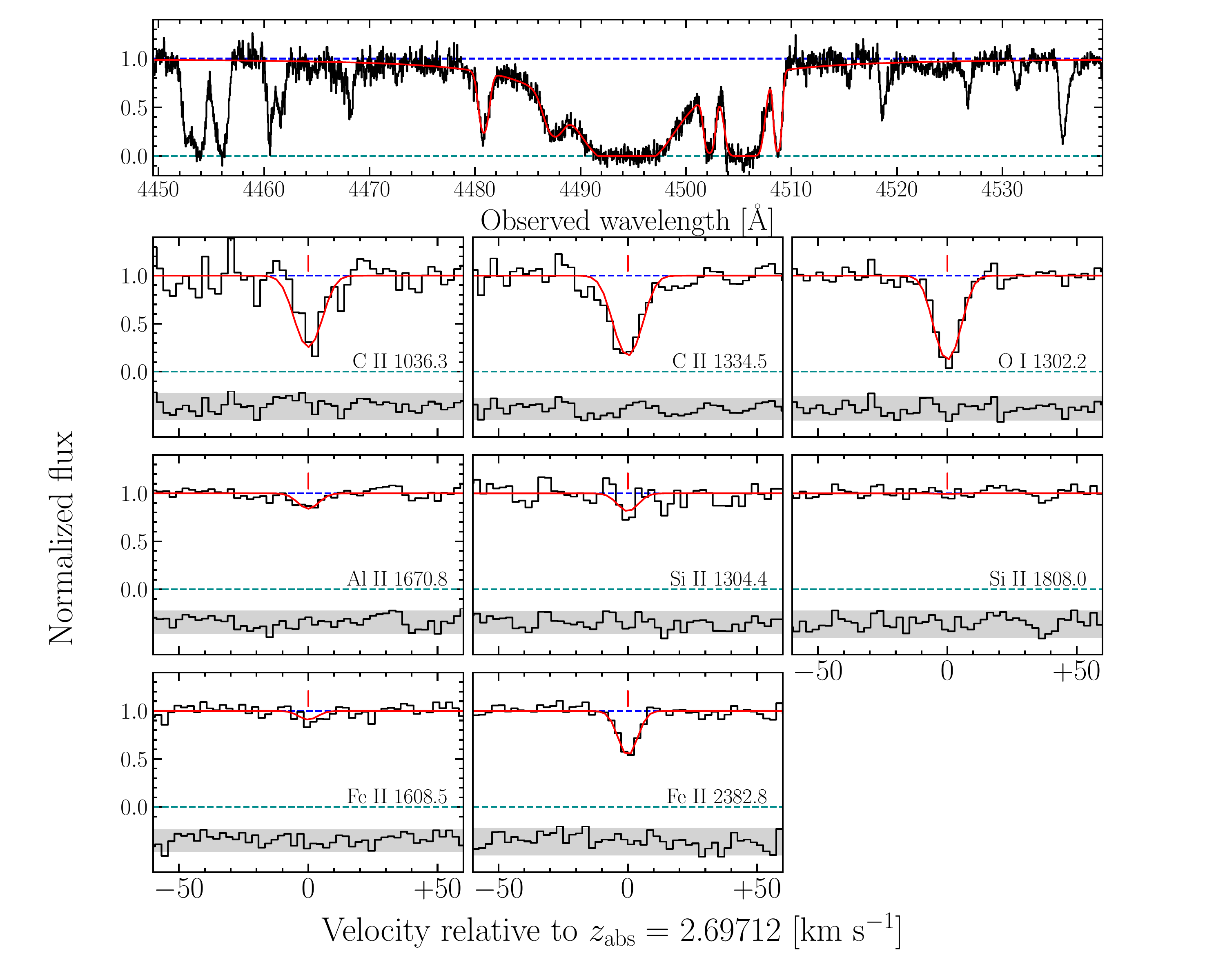}
    \caption{Same as Figure~\ref{fig:j0239} for the DLA at $z_{\rm abs}=2.69712$ towards J2150$+$0331. Again, note the information provided by the non-detection of Si\,{\sc ii} $\lambda1808$. }
    \label{fig:j2150}
\end{figure*}

\noindent  The absorber towards J2150$+$0331 is best modelled with one gaseous component at  $z_{\rm abs}=2.697125 \pm 0.000002$. The data, along with the best-fitting model are presented in Figure \ref{fig:j2150}, while the corresponding column densities are listed in Table \ref{tab:j2150}. \\
The broadening of this absorption line system is consistent with being entirely dominated by its thermal motions with a temperature of $T =  (1.6 \pm 0.4)\times10^{4}$~K. 
The narrow absorption features of this system combined with its metal paucity make it an ideal system for isotopic abundance studies with the ultrastable spectrograph ESPRESSO (Welsh et al. in prep.). 

We find that log$_{10}$ $N$(H\,{\sc i})/cm$^{-2}~= 19.68 \pm 0.05$. This is the lowest H\,{\sc i} column density system discovered by our programme so far. The column density is 4$\times$ lower than that of a bonafide DLA. 
We detect C\,{\sc iv}, as reported in Table~\ref{tab:j2150}, that is minimally offset from the central component. Based on the Fe\,{\sc ii} transitions $\lambda$1260, $\lambda$1608, and $\lambda$2382 and the H\,{\sc i} column density\footnote{We also cover the \fe{2344} feature however this is blended with telluric absorption.}, we infer an iron abundance [Fe/H]~$=-2.57\pm0.08$. 

\subsubsection{J2308$+$0854}
\begin{figure*}
    \centering
    \includegraphics[width=\textwidth]{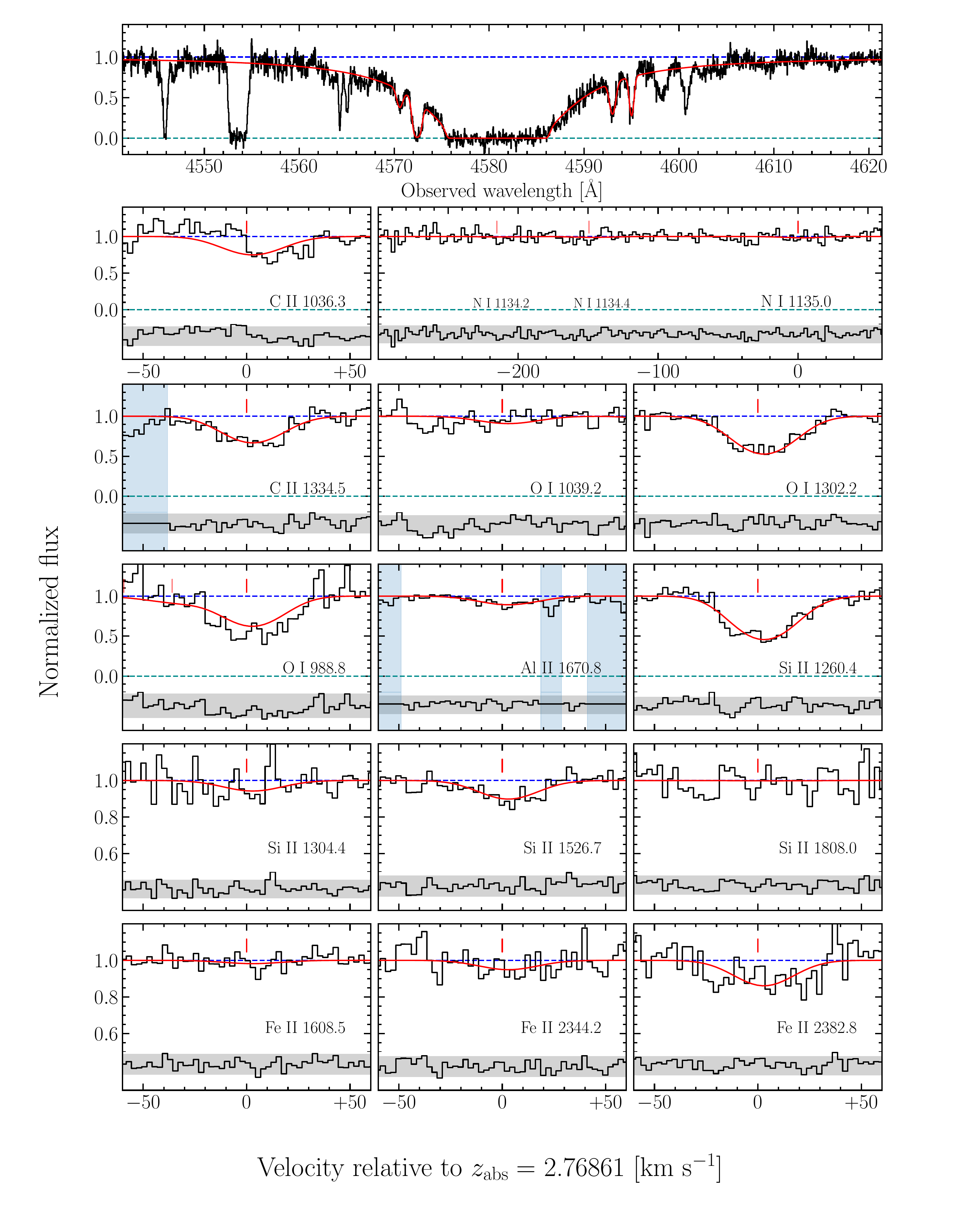}
    \caption{Same as Figure~\ref{fig:j0239} for the DLA at $z_{\rm abs}=2.76861$ towards J2308$+$0854.}
    \label{fig:j2308}
\end{figure*}

\noindent The absorber towards J2308$+$0854 is best modelled with one gaseous component at  $z_{\rm abs}=2.768607 \pm 0.000006$ with a turbulence of $b = 19 \pm 3$~\kms. During the analysis, we found that the total Doppler broadening of this gas cloud is dominated by turbulent motions. Thus, we adopt a temperature of $T=1.0\times10^{4}$~K, although, even with this temperature, the kinematics are completely dominated by the turbulent motions. 
The data, along with the best-fitting model are presented in Figure \ref{fig:j2308}, while the corresponding column densities are listed in Table \ref{tab:j2308}. We find that log$_{10}$ $N$(H\,{\sc i})/cm$^{-2}~= 19.95 \pm 0.05$ and [Fe/H]~$=-3.00 \pm 0.12$. It is a sub-DLA on the cusp of the EMP regime. It is additionally interesting because it is not typical to find such a broad system in this metallicity regime.

{\footnotesize \begin{table}
	\centering
	\caption{Same as Table \ref{tab:j0239} for the DLA at $z_{\rm abs}=2.76861$ towards the quasar J2308$+$0854. Again, note the distinct x-axis of the panel showcasing the N\,{\sc i} triplet and the non-detection of Si\,{\sc ii}~$\lambda 1808$.}
	\label{tab:j2308}
	\tabcolsep=0.10cm
\begin{tabular}{lccc}
\hline
Ion  & Transitions & $\log_{10} N$(X)/cm$^{-2}$ \\
  & used [\AA] &  &  \\ \hline \hline
H\,{\sc i}   &  1215                   & 19.95 $\pm$ 0.02  \\
C\,{\sc ii}  &  1036, 1334             & 13.50 $\pm$ 0.05   \\
N\,{\sc i}   &  1135                   & $\leq12.70^{\rm a}$ &  \\
O\,{\sc i}   &  1039, 1302             & 14.13 $\pm$ 0.03    \\
Al\,{\sc ii}   &  1670             & 11.72 $\pm$ 0.10    \\
Si\,{\sc ii} &  1190, 1260, 1304, 1526 & 12.84 $\pm$ 0.02   \\
Fe\,{\sc ii} &  1608, 2382       & 12.42 $\pm$ 0.10   \\ \hline
\end{tabular}
\begin{tabular}{l}
     $^{a}$ $3\sigma$ upper limit on column density. \\
\end{tabular}
\end{table} }

\subsection{Survey statistics}
\begin{figure}
    \centering
    \includegraphics[width=\columnwidth]{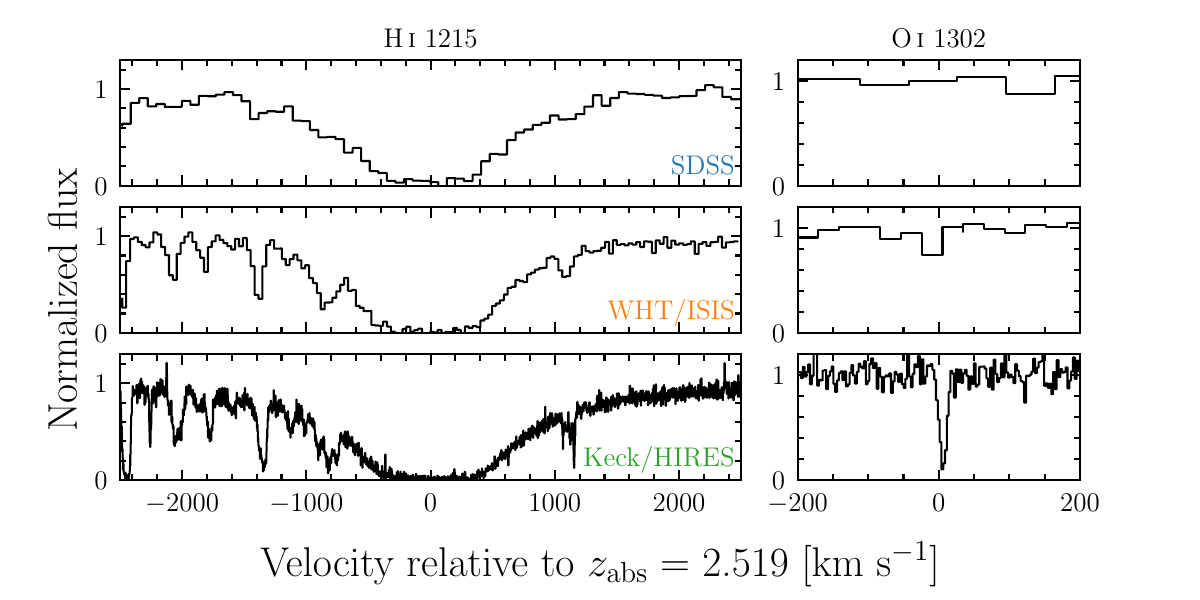}
    \caption{An example of one of the absorption line systems studied in this work observed with increasingly higher resolution instruments. Each row shows the H\,{\sc i} Ly$\alpha$ absorption and the region of spectrum where the strongest O\,{\sc i} metal line should appear. The first row shows the archival SDSS data analysed by \citet{Parks2018}. The middle row shows the data taken in 2019 with WHT/ISIS. The final row shows the data taken with the high resolution echelle spectrograph Keck/HIRES.}
    \label{fig:inc_res}
\end{figure}

Before discussing the detailed chemistry of our sample, we summarise the main findings of our search for metal-poor gas at cosmic noon. We have found four previously unknown high column density absorbers. All have log$_{10}$ $N$(H\,{\sc i})/cm$^{-2}~> 19.68$ and are at least [Fe/H]~$< -2.5$. One absorber, towards J1147$+$5034, is a bonafide EMP DLA with [Fe/H]~$=-3.13\pm0.06$. There is a further sub-DLA, towards J2150$+$0331 that falls on the borderline of this regime with [Fe/H]~$=-3.00\pm0.12$.  
Overall, 50 per cent of the gas clouds targeted with this programme can be considered EMP. The resulting metallicities of the discovered systems highlight the success of our technique employed to identify chemically unevolved environments at cosmic noon. Figure~\ref{fig:inc_res} highlights how high resolution echelle spectroscopy is ideally suited to detect the metal lines associated with such systems. 
We were originally focused on discovering bonafide DLAs. 
The measured H\,{\sc i} column densities are slightly lower than predicted based on the intermediate resolution data. Thus, they are most accurately described as sub-DLAs. Though, two-out-of-four of the newly discovered systems have column densities of log$_{10}$ $N$(H\,{\sc i})/cm$^{-2}~\gtrsim 20.0$, which is a regime where the resident ions are predominantly in a single ionization state.

An exceptional property shared across all absorbers discovered through this programme is the simple kinematic structures of the gas clouds. Out of the four literature EMP DLAs analysed in Section~\ref{sec:known}, only one can be modelled with one gaseous component\footnote{These four systems are those with previous high resolution data available.}. In contrast to this, all of these newly discovered systems (presented in Section~\ref{sec:new}) are accurately modelled with one gaseous component. In three-out-of-four cases, the turbulent broadening is found to be $b<5$~\kms. The relatively quiescent nature of these metal-poor absorbers is consistent with the relationship between metallicity and turbulent broadening that has been observed across the VMP DLA population \citep{CookePettiniJorgenson2015}, and indeed may be expected given the potential mass-metallicity relation associated with the DLA host galaxies \citep{Neeleman2013, Krogager2017}. In one case, the gas cloud can be entirely well-modelled via its thermal motions. It has recently been shown that this is an ideal environment to study isotopic abundances \citep{Welsh2020}. Indeed, while these absorbers are in the sub-DLA regime, the remarkably simple kinematic structures suggest that ionization corrections are likely to be negligible. 
\begin{table*}
        \caption{Metal abundances of our DLA sample}
    \label{tab:chem}
    	\tabcolsep=0.07cm
    \begin{tabular}{lccccccccc}
\hline
QSO & $z_{\rm abs}$ & $\log_{10}N($H\,{\sc i})/cm$^{-2}$ & [C/H] & [N/H] & [O/H] & [Al/H] & [Si/H] & [Fe/H] & [O/Fe] \\
\hline \hline
\multicolumn{9}{c}{previously known DLAs} \\
\hline

J0140$-$0839 & 3.697 &20.81 $\pm$ 0.05 & $-3.37 \pm 0.05$ & --- & $-2.92 \pm 0.05$ & $-3.56 \pm 0.06$ & $-3.13 \pm 0.05$ & $-3.37 \pm 0.06$ & $+0.45\pm 0.04$ \\

J0903$+$2628 & 3.077 &20.32 $\pm$ 0.05 & $-3.45 \pm 0.06$ & --- & $-3.08 \pm 0.05$ & $<-3.80$ & $-3.22 \pm 0.05$ & $<-3.65 $ & $>+0.57$ \\ 
J1001$+$0343 & 3.078 &20.20 $\pm$ 0.05 & $-3.06 \pm 0.05$ & $<-3.53$ & $-2.63 \pm 0.05$ & $-3.26 \pm 0.09$ & $-2.85 \pm 0.05$ & $-3.25 \pm 0.06$ & $+0.62\pm 0.05$\\ 

J1358$+$0349 & 2.853 &20.524 $\pm$ 0.006 & $-3.17 \pm 0.04$ & --- & $-2.85 \pm 0.04$ & $-2.94 \pm 0.02$ & $-2.87 \pm 0.02$ & $-3.45 \pm 0.04$ & $+0.60 \pm 0.06$ \\ 
\hline
\multicolumn{9}{c}{new DLAs} \\
\hline
 
J0239$-$0649 & 2.416 &20.10 $\pm$ 0.05 & $-2.44 \pm 0.09$ & $-3.33 \pm 0.16$$^{*}$ & $-2.07 \pm 0.10$ & $-2.73 \pm 0.13$ & $-2.21 \pm 0.07$ & $-2.64 \pm 0.06$ & $+0.57 \pm 0.10$ \\ 

J1147$+$5034 & 2.519 &20.23 $\pm$ 0.01 & $-3.06 \pm 0.03$ & --- & $-2.80 \pm 0.03$ & $-3.29 \pm 0.10$ & $-2.91 \pm 0.03$ & $-3.13 \pm 0.06$ & $+0.33 \pm 0.07$ \\ 
J2150$+$0331 & 2.697 &19.68 $\pm$ 0.05 & $-2.37 \pm 0.09$ & $-1.87 \pm 0.09$ & $-2.00 \pm 0.14$ & $-2.74 \pm 0.11$ & $-2.34 \pm 0.15$ & $-2.57 \pm 0.08$ & $+0.57 \pm 0.15$ \\ 
J2308$+$0854 & 2.769 &19.95 $\pm$ 0.05 & $-2.91 \pm 0.06$ & $<-3.07$ & $-2.51 \pm 0.06$ & $-2.66 \pm 0.11$ & $-2.62 \pm 0.05$ & $-3.00 \pm 0.12$ & $+0.49 \pm 0.11$ \\ 
\hline

\end{tabular}
\begin{tabular}{l}
$^{*}$ Note that this N detection is marginal.
\end{tabular}
\end{table*}

\begin{table}
	\centering
	\caption{Solar abundances adopted in this analysis from \citet{Asplund2021}. These are consistent with those recommended by \citet{Lodders2019} and also \citet{Asplund2009} and \citet{Magg2022} when considering the associated uncertainties.}
	\label{tab:solar_vals}
	\begin{tabular}{ccccc} 
		\hline
		Element & value \\
		\hline
		H & 12.00 \\
		C & 8.46  \\
		N & 7.83  \\
		O & 8.69  \\
        Al & 6.43  \\
        Si & 7.51  \\
        Fe & 7.46  \\
		\hline
	\end{tabular}
\end{table}

\section{Chemical abundances}

\label{sec:chem}
In this section, we present the chemical abundance patterns of near-pristine gas clouds at cosmic noon. The metal abundances are compiled in Table~\ref{tab:chem} assuming the \citet{Asplund2021} solar abundances that are provided in Table~\ref{tab:solar_vals}. Metal abundances are calculated based on the column densities of the components that are traced by O\,{\sc i} and Fe\,{\sc ii}. The relative column densities of the components associated with the strongest O\,{\sc i} absorption are the least likely to suffer from the effects of ionization; the components traced by neutral gas are more likely to have all their metals associated with a single dominant ionization stage. The abundances based on just these components are therefore more reliable than measuring the relative abundances from the total column densities of the detected ions. One of the primary goals of this survey is to test if EMP DLAs (those with ${\rm [Fe/H]}<-3.0$) exhibit an [O/Fe] abundance that is consistent with or different to VMP DLAs (those with $-3.0<{\rm [Fe/H]}<-2.0$).

\subsection{History of [O/Fe] at the lowest metallicities}
The behaviour of [O/Fe] at low metallicities has been the subject of many investigations in the field of chemical evolution \citep{McWilliam1997}. 
It is a key indicator of the onset of enrichment from Type Ia SNe. Indeed, the behaviour of [O/Fe] between $-2 <$~[Fe/H]~$<-1$ is well characterised for both the MW and the surrounding dwarf galaxies \citep[e.g.,][]{Matteucci1990, Kirby2011, Frebel2014, Bensby2017}. 
However, beyond [Fe/H]~$<-2$, determining O in stars is known to be challenging. Different diagnostic lines are often shown to produce conflicting results \citep{Garcia2006}.

The forbidden \of\ transition is considered to be a reliable tracer of O in stars since it is known to form in local thermodynamic equilibrium (LTE) \citep{Asplund2005}. This means the measurement of oxygen abundances using stellar spectra with these features can reach a higher degree of precision than those that require non-LTE corrections. Though, we note that the impact of 3D effects cannot be neglected and the associated correction must still be applied \citep{Nissen2002, Collet2007, Amarsi2019}. Critically, this forbidden feature is intrinsically weak and becomes challenging to detect at low metallicities. Thus, one requires high quality data in terms of both S/N and resolution to adequately detect this transition. Indeed, the reliable oxygen features in gas reservoirs suffer the opposite problem. Redwards of the Ly$\alpha$ forest there is the strong \oi{1302} feature. Given the large oscillator strength, this feature is often saturated when observing gas with H\,{\sc i} column densities approaching that of DLAs. 
This is naturally overcome when studying the lowest metallicity DLAs as the resulting absorption features of these systems are typically unsaturated. As a result, EMP DLAs are possibly the best environments to precisely and accurately assess the behaviour of [O/Fe] at the lowest metallicities. For an EMP DLA with [Fe/H]~$=-3.0$ and [O/Fe]~$=+0.40$, the corresponding column density of O\,{\sc i} would be $\log_{10}N($O\,{\sc i})/cm$^{2} = 14.4$. For a strong line like O\,{\sc i}~$\lambda 1302$, there is still a risk of saturation depending on the associated kinematics. In these cases, detecting unblended O\,{\sc i}~$\lambda 1039$ absorption would be invaluable. However, these systems are exceptionally rare and so it is challenging to determine the overall trends in the data. 

Past work has provided information on the behaviour of [O/Fe] in the VMP regime \citep{Cooke2011b}. This work indicated that both VMP DLAs and VMP stars exhibit a similar distribution of [O/Fe] abundances and could be accurately modelled with a typical [O/Fe] abundance of [$\langle$O/Fe$\rangle$]~$\simeq +0.40$. At the time, there was tentative evidence that this abundance ratio would increase within the EMP regime. Our latest data provide a marked improvement in both the quality and quantity of data available for the \emph{most} metal-poor DLAs. 

\subsection{The behaviour of [O/Fe] at cosmic noon}
We have completed our investigation of the [O/Fe] abundance of known EMP DLAs. In all cases we have improved the precision of the [O/Fe] abundance determinations by at least a factor of 3 or, in the case of J0903+2628, improved the upper limit by $\sim$1 dex. These improved measures are combined with our sample of [O/Fe] abundances of the newly discovered absorbers. We now use these data to reassess the behaviour of [O/Fe] at the lowest metallicities. 

\begin{figure*}
    \centering
    \includegraphics[width=\textwidth]{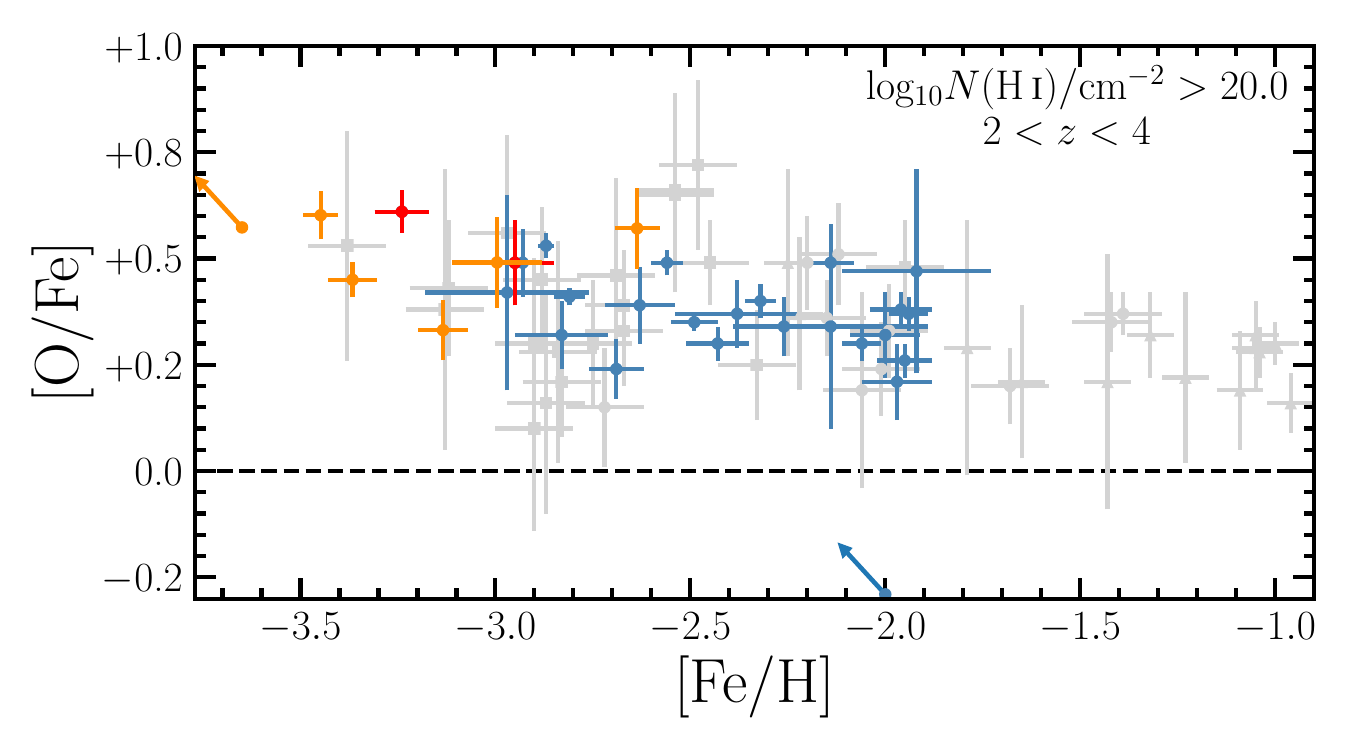}
    \caption{[O/Fe] vs [Fe/H] of DLAs and sub-DLAs. Blue data are from the literature compilation in \citet{Welsh2022} and the red points are the data presented in that work. The orange points are the DLAs presented in this work combined with the results of \citet{Welsh2023}. The arrows represent systems with Fe\,{\sc ii} column densities known as upper limits. Note that we do not plot the abundances of J2150$+$0331 in this figure since the column density of neutral hydrogen does not meet the criteria of $\log_{10}N($H\,{\sc i})/cm$^{-2}>20.0$ (within the associated errors). The abundances of metal-poor stars are shown as grey symbols. }
    \label{fig:ofe_new}
\end{figure*}

Figure~\ref{fig:ofe_new} shows the [O/Fe] abundance ratios as a function of the Fe-metallicity updated to include the DLAs analysed and presented in this work (orange symbols), together with our \citet{Welsh2022} literature compilation of VMP and EMP DLAs (blue symbols), and MW halo stars \citep[grey symbols;][]{Nissen2002, Cayrel2004, Garcia2006}. In this plot, we only consider systems that have an H\,{\sc i} column density in excess of $\log_{10}N($H\,{\sc i})/cm$^{-2}>20.0$. 
Note that the [O/Fe] abundances of these MW halo stars shown in this plot have been calculated using the [O\,{\sc i}] 6300\AA\ absorption feature. They have further been corrected for 3D effects as discussed in \citet{Cooke2011b}.

\begin{figure*}
    \centering
    \includegraphics[width=\textwidth]{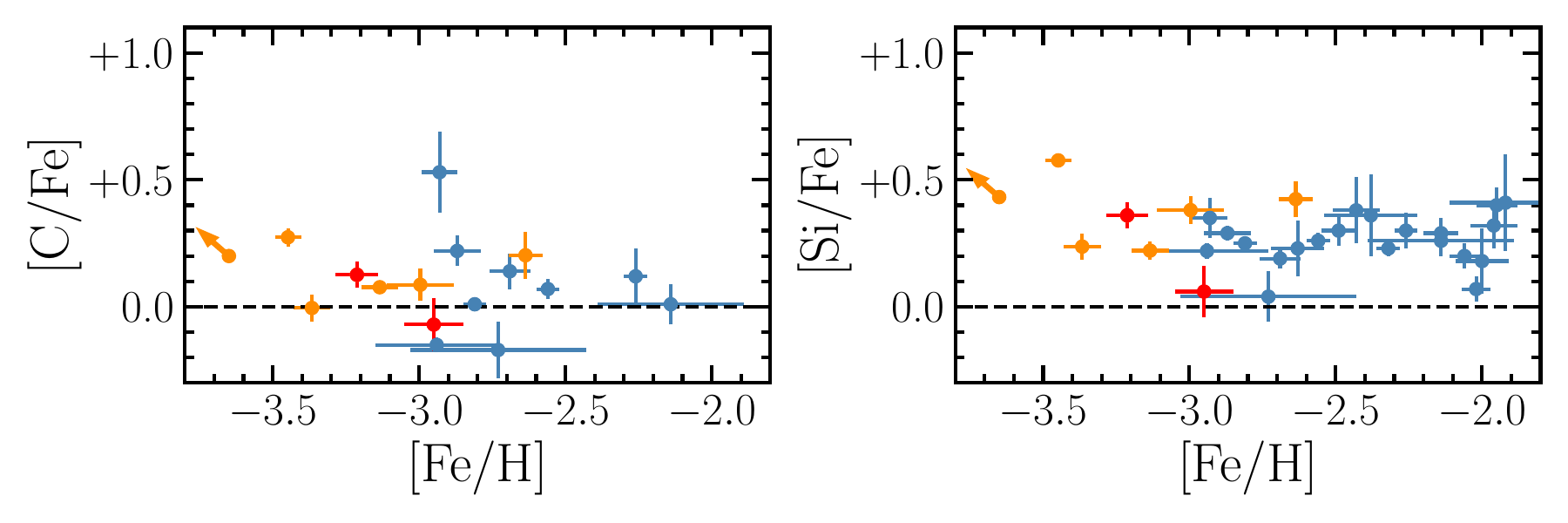}
    \caption{Same as Figure~\ref{fig:ofe_new} but the panels showcase [C/Fe] and [Si/Fe] respectively.
    \label{fig:cfe}}
\end{figure*}

Overall, the full sample of metal-poor absorbers provide increased statistics to assess the behaviour of [O/Fe] as a function of Fe-metallicity. There are just three measurements of [O/Fe] in stars when [Fe/H]~$<-3.0$ compared to the six DLAs reported here. This highlights the important role of new [O/Fe] measurements in this metallicity regime. 

 We also consider the Fe-evolution of [C/Fe] and [Si/Fe]. Figure~\ref{fig:cfe} shows the [C/Fe] and [Si/Fe] abundance ratios as a function of their Fe-metallicity.

\subsection{Intrinsic scatter}

\begin{figure*}
    \includegraphics[width=\textwidth]{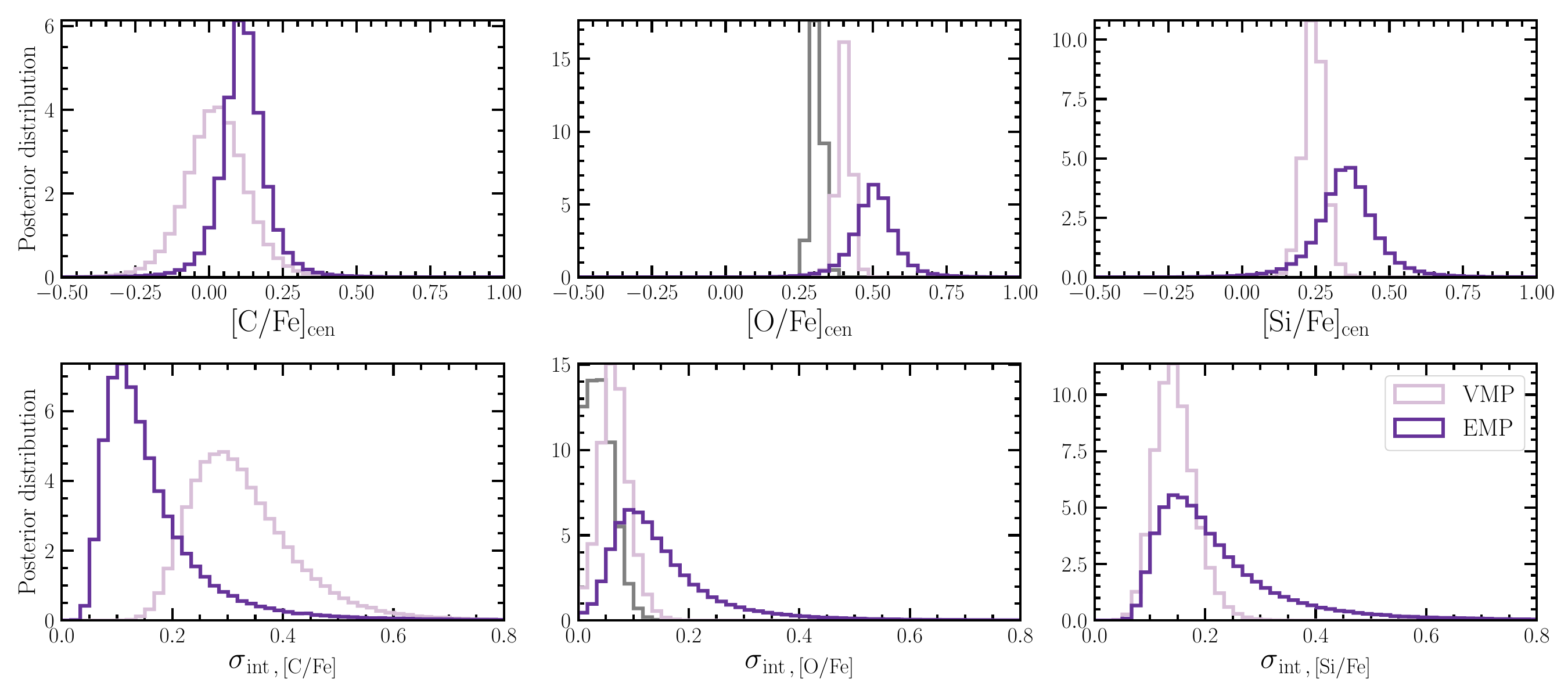}
    \caption{\label{fig:intsca} Results of the MCMC analysis investigating the intrinsic scatter in the observed data across the VMP (pink) and EMP (purple) regime. The top row shows the posterior distributions of the central [X/Fe] values in order of their atomic number (C, O, and Si). The bottom row shows the associated distributions of the intrinsic scatter. In the middle column we show in grey the same analysis of the intrinsic scatter associated with the VMP stellar data included in Figure~\ref{fig:ofe_new}. 
    }
\end{figure*}

To quantify the behaviour seen in Figure~\ref{fig:ofe_new} and Figure~\ref{fig:cfe}, we investigate the intrinsic scatter present in the data across the VMP and EMP regime. First, we categorise these absorbers based on the Fe-metallicity. We then assume that the observed abundance ratios in each regime can be well-modelled with a Gaussian centred at a given [X/Fe] value ([X/Fe]$_{\rm cen}$) and an associated intrinsic scatter $\sigma_{\rm int \, ,\, [X/Fe]}$. The total dispersion of the sample is therefore given by the measurement error and the intrinsic scatter added in quadrature. Abundance ratios that are only known as either upper or lower limits are excluded from consideration. To find the best-fitting model given these VMP and EMP data, we run a Markov Chain Monte Carlo (MCMC) likelihood analysis using the {\sc emcee} software package \citep{EMCEE}. Figure \ref{fig:intsca} shows the resulting posterior distributions of the investigated parameters for [C/Fe], [O/Fe], and [Si/Fe]. In all cases, these results highlight the elevated [X/Fe] ratios observed in the EMP regime compared to the VMP regime.

For C, we find [C/Fe]$_{\rm cen} =+0.02 \pm 0.07$ and $\sigma_{\rm int \, ,\, [C/Fe]} = 0.31_{-0.05}^{+0.07}$ for VMP DLAs. Here, and subsequently, we report the median value of the distributions while the errors are given by the associated interquartile range. Comparatively, the [C/Fe] abundance ratio of EMP DLAs can modelled with [C/Fe]$_{\rm cen} =+0.11 \pm 0.04$ and $\sigma_{\rm int \, ,\, [C/Fe]} = 0.14_{-0.04}^{+0.06}$. 
Interestingly, the intrinsic scatter of the EMP [C/Fe] ratios is comparatively less than that observed for the VMP data. This could, in part, be due to the quality of the data available for C in the VMP regime. C is typically measured using the strong C\,{\sc ii} $\lambda$1334 line. When available the C\,{\sc ii} $\lambda$1036 line can also be used but it is often blended with features in the Ly$\alpha$ forest. At these high H\,{\sc i} column densities the accessible C\,{\sc ii} features easily become saturated.

For O, we find [O/Fe]$_{\rm cen} =+0.40 \pm 0.02$ and $\sigma_{\rm int \, ,\, [O/Fe]} = 0.06 \pm 0.02$ for VMP DLAs. The O abundance of EMP DLAs can modelled with [O/Fe]$_{\rm cen} =+0.50 \pm 0.04$ and $\sigma_{\rm int \, ,\, [O/Fe]} = 0.13_{-0.04}^{+0.06}$. Both the central value and associated intrinsic scatter are found to be distinct across these metallicity regimes. 
This increased intrinsic scatter in the distribution of the [O/Fe] abundance ratios in the EMP regime compared to the VMP regime is reminiscent of the behaviour predicted in \citet[their Fig. 2]{Welsh2019} of an increased dispersion across the abundance ratios caused by a fewer number of enriching stars. The ability to investigate the intrinsic scatter is dependent on both the size of the sample and the relative precision of the abundance ratios. Despite the small size of the EMP DLA sample, given the errors associated with the intrinsic scatter determinations, these data appear sufficient to produce similarly competitive constraints when compared to intrinsic scatter found using the larger sample of VMP DLAs. For reference we also investigate the intrinsic scatter associated with the VMP stellar [O/Fe] data shown in Figure~\ref{fig:ofe_new}. Interestingly, we find that, unlike the VMP DLAs, the intrinsic scatter in these data are consistent with minimal excess dispersion --- [O/Fe]$_{\rm cen} =+0.31 \pm 0.01$ and $\sigma_{\rm int \, ,\, [O/Fe]} = 0.04 \pm 0.02$. At present, it is unclear if the excess dispersion seen in VMP DLAs is due to underestimated errors, or if DLAs exhibit a real intrinsic dispersion relative to stars of a comparable metallicity. The latter possibility could indicate that VMP DLAs are not as well-mixed as the birth sites of VMP stars. Furthermore, the VMP DLA data represent the abundances of different structures (e.g., galaxies). These different galaxies may have experienced different star formation histories, perhaps leading naturally to a higher degree of scatter in the [O/Fe] ratio in DLAs compared to that of MW halo stars. A bespoke programme tailored to understanding the precise chemistry of VMP DLAs would address this possibility.

The behaviour of [Si/Fe] is similar to that of [O/Fe]. For Si, we find [Si/Fe]$_{\rm cen} =+0.24 \pm 0.02$ and $\sigma_{\rm int \, ,\, [Si/Fe]} = 0.14\pm 0.03$ for VMP DLAs.  Comparatively, the Si abundance of EMP DLAs can modelled with [Si/Fe]$_{\rm cen} =+0.35 \pm 0.06$ and $\sigma_{\rm int \, ,\, [Si/Fe]} = 0.19_{-0.05}^{+0.08}$.  In this case, the intrinsic scatter of [Si/Fe] seems to peak at a similar position across both the VMP and EMP regime. However, as is the case for [O/Fe], the tail of the scatter extends to larger values for the EMP data.

Finally, we note that the [O/Fe]$_{\rm cen}$ and $\sigma_{\rm int \, , \, [O/Fe]}$ values found via this analysis for both the VMP and EMP sample are consistent with our previous results \citep{Cooke2011b, Welsh2022}. 

\begin{figure*}
    \includegraphics[width=\textwidth]{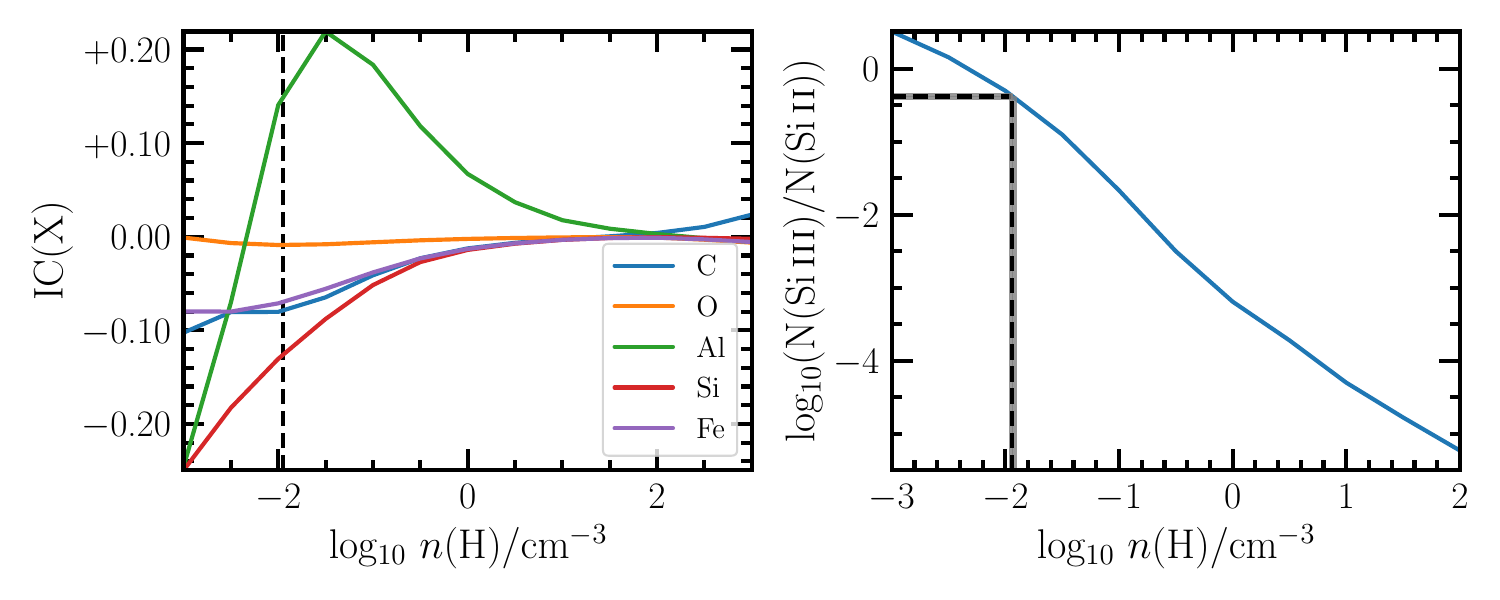}
    \caption{Left: Ionization corrections for the EMP DLA towards J1147+5034. We plot the expected correction for all of the observed elements as a function of gas volume density. The vertical dashed line correspond to the estimated gas density of the DLA. Right: The column density ratio of successive ion stages of Si versus gas density. We also plot the observed values of the Si\,{\sc iii}/Si\,{\sc ii} ion ratio (dashed lines with grey shaded region encompassing the associated errors) and use this to determine the associated gas density. This highlights that O\,{\sc i} is stable against ionization corrections and the impact on Fe\,{\sc ii} would only serve to increase the enhancement of [O/Fe] observed for EMP DLAs relative to VMP DLAs.
    \label{fig:ic}}
\end{figure*}

\section{Abundance corrections}
\label{sec:phys}
Neutral gas reservoirs are some of the best astrophysical environments to determine the chemical abundance patterns of the most abundant chemical elements at high precision. However, there are a number of physical processes that must be considered when calculating the metal abundances from the measured gas-phase column densities. The two main physical processes that can impact abundance determinations are ionization effects and dust depletion.

\subsection{Ionization corrections}
\label{sec:ioncorr}
Given the column densities of neutral hydrogen recorded for some of the newly discovered gas clouds in our sample, they are more appropriately described as sub-DLAs rather than bonafide DLAs. Gas clouds with an H\,{\sc i} column density less than the DLA threshold may not be able to self-shield as effectively as DLAs. 
Thus, we may need to apply some ionization corrections to the observed ion column densities to uncover the intrinsic metal abundances. To minimise the impact of ionization effects, we have solely considered the relative abundances of the components traced by neutral oxygen. This restricts our analysis to the predominantly neutral gas. Moreover, we find no evidence for a statistically significant correlation between the relative ion column densities and the H\,{\sc i} column density across our sample. This suggests that the sample sub-DLAs are not empirically different to the bonafide DLAs.

We can further test the appropriateness of this choice using the ion column densities of elements that have been detected in multiple ionization states. In the case of J1147$+$5034, we have information on the N(Si\,{\sc ii})/N(Si\,{\sc iii}) ratio. The abundances of successive ions of an element can be used, alongside photoionization models, to estimate the ionization correction. We then assume that this ionization correction is appropriate for the other sub-DLAs in our sample. We define this small correction to be the difference between the true intrinsic abundance and the measured abundance based on the column densities of the dominant ionization stages:
\begin{equation}
    \label{eqn:ic}
    {\rm [X/H] = [X_{N}/H\,\textsc{i}] + IC(X)   }
\end{equation}
for each observed element, X. To estimate the magnitude of such corrections, we used the \textsc{cloudy} photoionization software developed by \citet{Ferland1998, Ferland2017} to model the newly discovered EMP DLA towards J1147$+$5034. We model the DLA as a plane-parallel slab of constant volume density gas in the range $-3< {\rm log}_{10}~n({\rm H)/cm}^{-3} <~3$, irradiated by the UV background as described in \citet{Haardt2012} at $z_{\rm abs}=2.588$. At this epoch, the latest determination of the UVB from \citet{Khaire2019} is consistent with that of \citet{Haardt2012}. Using the solar abundance scale in \citet{Asplund2021}, we then scaled the metal abundances to that of the DLA ([Fe/H]~$=-3.13 \pm 0.06$). The simulations were stopped once the H\,{\sc i} column density of the DLA was reached, at which point we output the ion column densities of the slab. As will be discussed in Section~\ref{sec:dust}, we assume that the impact of dust at these metallicities is negligible and is therefore not included in the model.  Given the assumed background radiation field, the ionization correction of each element depends on the volume density of the gas. The resulting ionization corrections calculated as a function of the gas volume density are show in the left-hand panel of Figure~\ref{fig:ic} while the associated ion ratio of the stages of Si are shown in the right-hand panel of Figure~\ref{fig:ic}. These results are consistent with the alternative approach of scaling the abundances to that of a typical VMP DLA \citep[from][]{Cooke2011b}. Based on these calculations, the gas volume density of the DLA J1147$+$5034 can be estimated by considering the ratio of the successive ion stages of Si. The column densities in Table~\ref{tab:j1147} indicate that, for this DLA, there is 2.4 times more Si\,{\sc ii} than Si\,{\sc iii}. This implies a gas density of ${\rm log}_{10}~n({\rm H)/cm}^{-3} \simeq -1.95$.  Thus, the left-hand panel of Figure~\ref{fig:ic} indicates that the ionization corrections for all observed elements are $<0.15$~dex. Indeed, the negative ionization correction associated with Fe\,{\sc ii} indicates that the intrinsic [Fe/H] abundance is \emph{lower} than that measured. This would place the DLA further into the EMP regime and enhance the elevated [O/Fe] abundance observed. Interestingly, a similar trend is seen for the ionization correction associated with C\,{\sc ii}. This suggests that the intrinsic [C/H] abundance is lower than that measured. Though, the observed [C/Fe] may remain unchanged given the similar correction expected for both ions. This cannot be said for Si\,{\sc ii}. The negative ionization correction associated with Si\,{\sc ii} is larger than that associated with Fe\,{\sc ii}. This could potentially cancel out any differences observed between the [Si/Fe] of VMP and EMP DLAs (Fig.~\ref{fig:intsca}; top right panel). Note that irrespective of the volume density, the ionization correction expected for [O/H] is negligible.  

\subsection{Dust depletion}
\label{sec:dust}
The impact of dust depletion is believed to be negligible when the iron abundance is [Fe/H]~$<-2$ \citep{decia2018, Vladilo2018}. The depletion of Si in particular is expected to be negligible for these low metallicity DLAs \citep{Vladilo2011}. Recent progress has been made towards an improved understanding about the effects of dust depletion in low metallicity environments, thanks to new measurements of dust depletion in the Large Magellanic Cloud \citep[LMC;][]{Romanduval2022} and low metallicity dwarf galaxies \citep{Hamanowicz2024}. While these studies do not reach the EMP regime studied in this paper, they do confirm that the trend of decreasing dust depletion with decreasing metallicity persists in all currently investigated environments. Indeed, if dust depletion were to be considered, it would only serve to emphasise the trend shown in Figure~\ref{fig:ofe_new}. In this scenario the VMP DLAs would in fact show lower [O/Fe] ratios and therefore the difference between the mean abundance across the VMP and EMP regimes would be increased. An empirical test of dust depletion at these metallicities would be possible if one were to obtain a measurement of ions like Mg\,{\sc ii} or Zn\,{\sc ii}; while challenging with current facilities, these measurements will become possible with the next generation of ground-based telescopes.

\subsection{Redshift evolution of [C/O]}

\begin{figure}
    \centering
    \includegraphics[width=\columnwidth]{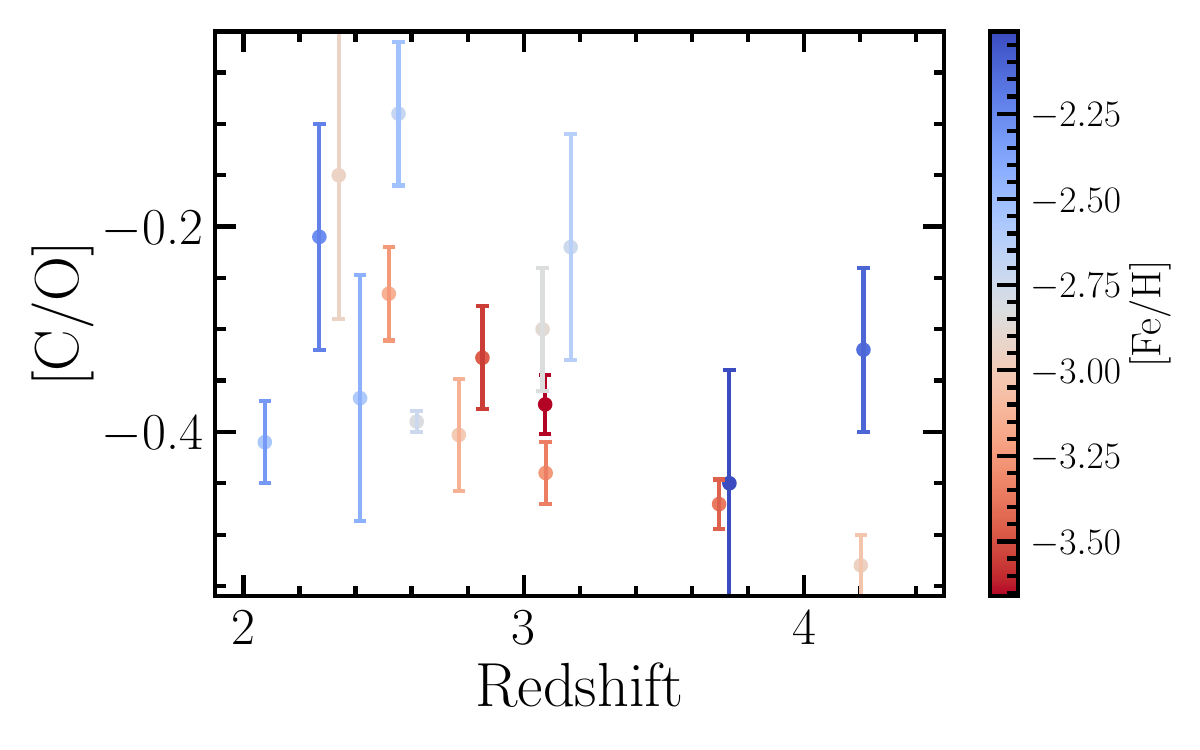}
    \caption{[C/O] ratio of VMP/EMP DLAs and sub-DLAs as function of the absorption redshift. We colour the systems by metallicity (see colour bar). The \emph{most} metal-poor systems (i.e, within 0.1~dex of the EMP regime) are indicated by the red points. These data show a trend of increasing [C/O] at lower redshifts that is not seen across the more metal rich absorbers. This indicates that these most metal-poor systems may be showcasing a unique signature of reionization quenching. This is in line with the expectation that only the most metal-poor systems would be impacted by this process at this epoch.}
    \label{fig:co}
\end{figure}

The star formation histories of low mass structures at cosmic noon are encoded in the chemistry of the most metal-poor DLAs. In \citet{Welsh2020}, we presented tentative evidence of a redshift evolution of the [C/O] ratio of the most metal-poor DLAs. 
From chemical evolution models, one may expect to witness an increase in the [C/O] ratio at late times due to the onset of enrichment from the intermediate and low-mass stars that are bound within the same host of these gas clouds \citep{Akerman2004, Cescutti2009, Romano2010}. The low and intermediate mass stars yield a significant amount of carbon near the end of their lives, particularly the asymptotic giant branch (AGB) stars that have masses $M\sim(2 -7)$~M$_{\odot}$ \citep{Karakas2007, Karakas2010}. Moreover, the carbon yields of low- and intermediate-mass stars are thought to increase with decreasing metallicity \citep[][]{Chiappini2003}. Thus, given the metallicities of these DLAs, any associated enrichment from longer-lived stars could introduce substantial amounts of carbon. This is the case for both the (hypothetical) low-mass Pop III stars \citep[][]{CampbellLattanzio2008} and the intermediate-mass Pop II stars \citep[e.g.][]{Karakas2010}.  

Given the redshifts of these newly discovered systems, this sample is ideally placed to investigate the relationship between [C/O] and redshift. This sample contains both the lowest redshift EMP DLA absorber discovered to date and new data on the highest redshift EMP DLA discovered to date. 
 We plot the [C/O] abundance ratio as a function of redshift for these latest data alongside the literature sample in Figure~{\ref{fig:co}}. In this case, we color the data points based on their Fe-metallicity. 
This figure highlights that the lowest metallicity systems (i.e. those with [Fe/H] within 0.1 dex of the EMP regime) support this trend of increasing [C/O] as redshift decreases. It is \emph{only} in the \emph{most} metal-poor objects that one might expect to uncover such a trend, since these systems would have experienced less ongoing star formation. Objects that have experienced ongoing star-formation are more likely to be more continuously enriched with the products of massive stars, and thus the yields of any associated AGB stars would be spread over a longer period of time. Indeed, the relatively delayed onset of enrichment from the intermediate mass stars at $z\sim3$, witnessed through the increasing [C/O] abundance at later times, may be an indication that these most metal-poor systems experience a period of quenched star-formation after the epoch of reionization.

\section{Discussion}
\label{sec:disc}
Our observational campaigns have yielded a high success rate of finding EMP DLAs. Of the 9 systems observed with WHT/ISIS that had sufficient final S/N ratios, 5 of these were selected for follow-up observations with either UVES or HIRES, and 2 of them are confirmed EMP DLAs or EMP sub-DLAs (with one further candidate awaiting observations). All newly discovered absorbers have an iron abundance [Fe/H]~$<-2.5$. This has allowed us to empirically investigate the chemical evolution of [O/Fe] across the most metal-poor systems with a larger sample size than previously available. 
Beyond empirical trends at cosmic noon, we can also compare our results to chemical evolution models to infer the properties of the enriching stars, and the most likely enrichment histories of near-pristine gas.

\subsection{Stochastic enrichment of near-pristine DLAs}
\label{sec:model}
 Using the observed abundance pattern of these DLAs, alongside the stochastic chemical enrichment model first described by \citet{Welsh2019}, we can investigate the possible enrichment history of near-pristine gas at cosmic noon. 

In this model, the mass distribution of Population III stars is modelled as a power-law: $\xi(M)=k\,M^{-\alpha}$, where $\alpha$ is the power-law slope\footnote{In our formulation, $\alpha=2.35$ corresponds to a Salpeter IMF.}, and $k$ is a multiplicative constant that is set by defining the number of stars, $N_{\star}$, that contribute to the enrichment between a minimum mass $M_{\rm min}$ and maximum mass $M_{\rm max}$, given by:
\begin{equation}
\label{eqn:N}
\centering
    N_{\star} = \int_{M_{\rm min}}^{M_{\rm max}} kM^{-\alpha} dM  \; .
\end{equation}
We note that $M_{\rm max}$ represents the maximum \emph{enriching} mass; stars of higher mass can form, but in our model, these stars are assumed to directly collapse to black holes and do not contribute to the enrichment \citep[see][]{Sukhbold2016}. Since the first stars are thought to form in small multiples, this underlying mass distribution is necessarily stochastically sampled. We utilise the yields from simulations of stellar evolution to construct the expected distribution of chemical abundances given an underlying IMF model. These distributions can then be used to assess the likelihood of the observed DLA abundances given an enrichment model.

We utilise the yields from \citet{HegerWoosley2010} (hereafter \citetalias{HegerWoosley2010}) to construct the expected distribution of chemical abundances given an underlying IMF model. The \citetalias{HegerWoosley2010} simulations model Population III progenitors with an initial metallicity Z=0. The simulations trace nucleosynthesis processes during the lifetime of the star as well as the explosive nucleosynthesis produced by the core-collapse SNe (CCSNe). The calculated yields reported by \citetalias{HegerWoosley2010} represent the mass of each metal that is returned to the interstellar medium.  These simulations explore initial progenitor masses in the range $M=(10-100)~{\rm M_{\odot}} $, explosion energies from $E_{\rm exp}=(0.3-10)\times10^{51}$ erg, and mixing parameters from $f_{\rm He} = 0 - 0.25$. The explosion energy is a measure of the final kinetic energy of the ejecta at infinity, while the mixing between stellar layers is parameterised as a fraction of the helium core size. This parameter space is evaluated across 120 masses, 10 explosion energies, and 14 mixing parameters. We linearly interpolate between this grid of yields during our analysis. 
We refer the reader to \citetalias{HegerWoosley2010} and \citet{Welsh2019} for further details and caveats related to these yield calculations. 

 Our model contains six parameters: $N_{\star}$, $\alpha$, $M_{\rm min}$, $M_{\rm max}$, $E_{\rm exp}$ and $f_{\rm He}$. The range of model parameters we consider are:
\begin{eqnarray*}
-5 \leq &\alpha& \leq  5 \; , \\ 
 1 \leq &N_{\star}& \leq 150 \; , \\
 0.3 \leq &E_{\rm exp}/10^{51}{\rm erg}& \leq 10  \; ,\\
 20 \leq & M_{\rm max}/ {\rm M_{\odot}}& \leq 70 \; .
\end{eqnarray*}
In what follows, we assume that stars with masses $>10~{\rm M_{\odot}}$ are physically capable of undergoing core-collapse. Therefore, we fix $M_{\rm min}=10~{\rm M_{\odot}}$. Since we are only considering the yields of CCSNe, we are not sensitive to the enrichment from lower mass stars. We also assume that the mixing during the explosive burning phase of the SN can be well parameterised by a fraction of He core size. Specifically, the mixing between the stellar layers is smoothed using a boxcar filter that is $10$~per cent of the helium core size (i.e., $f_{\rm He}=0.100$). This is the fiducial choice from \citetalias{HegerWoosley2010} based on their comparison with the \citet{Cayrel2004} stellar sample. These empirically motivated constraints reduce the number of free parameters to 4. We apply uniform priors to all remaining model parameters. Note that the upper bound on $M_{\rm max}$ corresponds to the mass limit above which pulsational pair-instability SNe are believed to occur \citep{Woosley2017}.

\begin{figure*}
    \centering
    \includegraphics[width=.7\textwidth]{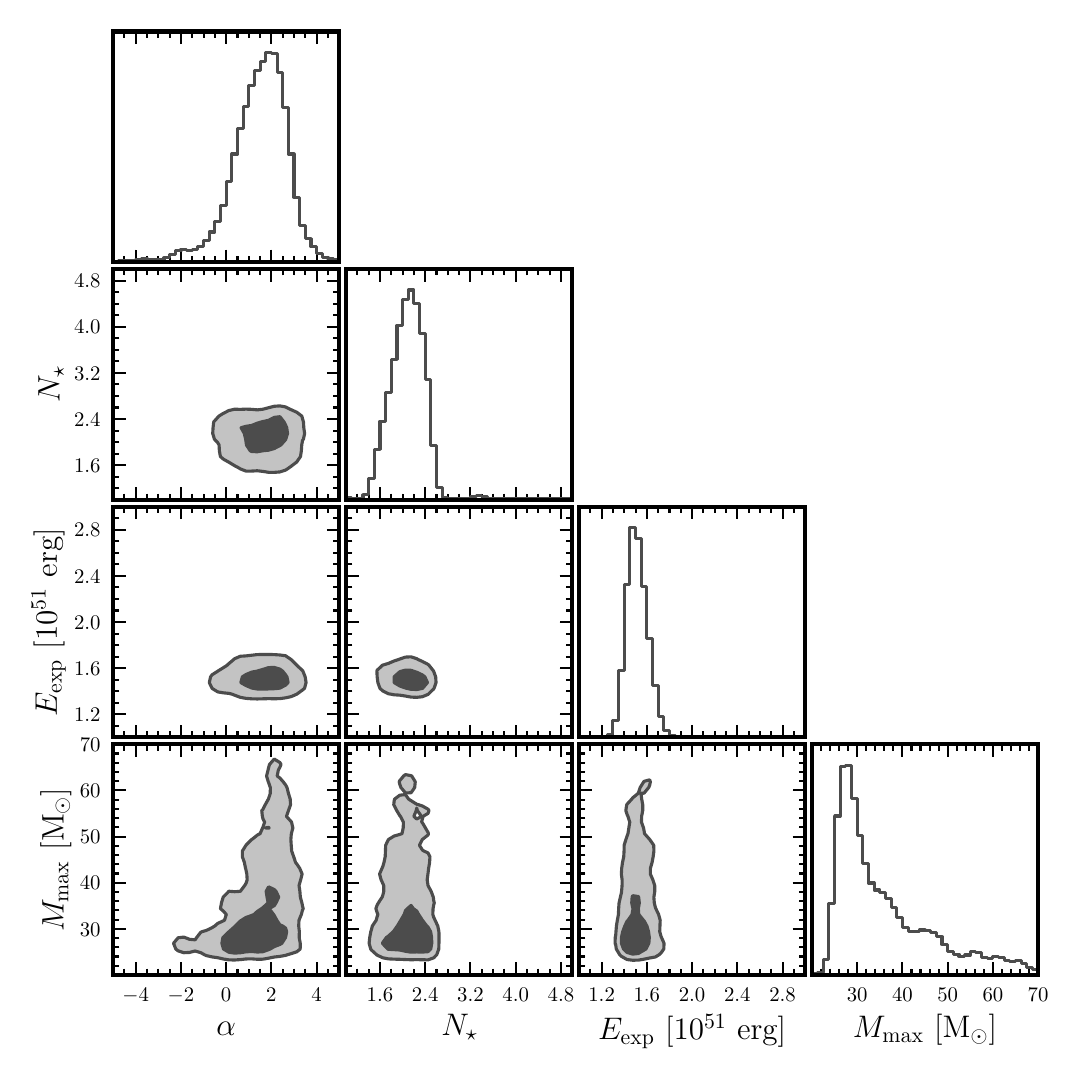}
    \caption{The marginalised maximum likelihood distributions of our fiducial model parameters (main diagonal), and their associated 2D projections, given the measured chemistry of the EMP DLAs in this sample. The dark and light contours show the $68\%$ and $95\%$ confidence regions of these projections respectively. The EMP DLAs in this sample can be well modelled via the enrichment from a small handful of progenitors that ended their lives as low energy CCSNe. }
    \label{fig:mcmc_res}
\end{figure*}

\subsection{Application to EMP DLAs}
Using the [C/O], [Al/O], [Si/O], and [Fe/O] abundance ratios of the EMP DLAs in this sample, we run a MCMC maximum likelihood analysis, again using the {\sc emcee} software package, to find the enrichment model that best fits these data. The converged chains are shown in Figure~\ref{fig:mcmc_res}. 

\subsubsection{IMF slope}
The slope of the IMF is found to be consistent with a Salpeter distribution, however it shows a slight preference towards a more top-heavy IMF slope. The distribution can be described by $\alpha = 1.57_{-0.77}^{+0.69}$ where here, and subsequently, the reported parameter values are the median values and the interquartile range of the parameter distributions. 
\subsubsection{Number of enriching stars}
The inferred number of enriching stars shows a preference towards low values of $N_{\star}$. The distribution is well-centred on $N_{\star} = 2\pm 1~$ (when rounded to integer values). This is markedly lower than the number of enriching stars found for a more typical metal-poor ([Fe/H]~$\sim-2.5$) DLA. In \citet{Welsh2019}, we found the number of enriching stars to be $N_{\star} < 72~(2\sigma)$ with a maximum likelihood value of $N_{\star}\sim10$. These results indicate that the typical number of enriching stars decreases when considering the lowest Fe-metallicity DLAs. We have repeated the analysis of \citet{Welsh2019} after removing the known EMP DLAs and find the resulting distributions are consistent with the above scenario. The inferred $N_{\star}$ when  exclusively considering VMP DLAs is higher than that found for EMP DLAs. Indeed, uniquely in the EMP regime, we are approaching a typical number of enriching stars that may be expected from one Pop. III minihalo \citep{Skinner2020}. 

\subsubsection{Typical explosion energy}
The distribution of the typical explosion energy is well-centred around $E_{\exp}=(1.6\pm 0.6)\times10^{51}$~erg. This is considered a moderate energy explosion for a CCSN. Indeed, this result is also in line with the carbon-enhanced absorbers that have been detected at cosmic noon \citep{Zou2020, Saccardi2023, Salvadori2023}. One way to explain this relative enhancement of [C/Fe] is via enrichment from stars with low energy explosions \citep{Umeda2003}. These SNe are rich in light atomic number elements (like C) since they are more easily expelled during the explosion process.

\begin{figure*}
    \centering
    \includegraphics[width=.9\textwidth]{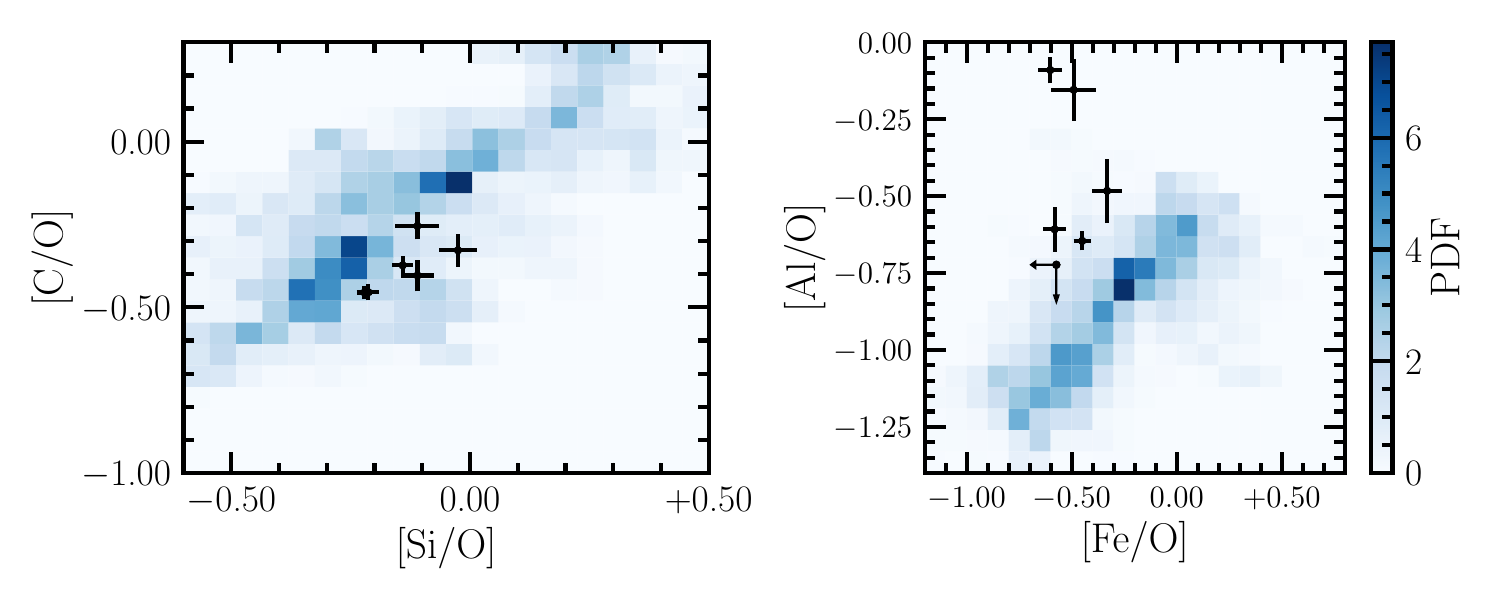}
    \caption{The [C/O], [Si/O], [Al/O], and [Fe/O] abundance ratios of the systems that make up our sample (black symbols with error bars) overplotted on the joint probability distributions (blue shaded distributions) of [C/O] and [Si/O] (left) and [Al/O] and [Fe/O] (right) of the best-fit enrichment model. Overall the EMP DLA data are well-described by the inferred enrichment model. Though, there are signs that the DLAs with [Al/O] abundances close to the solar value may be challenging to reproduce.}
    \label{fig:new_data_w_popiii_model}
\end{figure*}

\subsubsection{Maximum enriching mass}
The distribution of the maximum enriching mass is well-centred around $M_{\rm max}=32_{-4}^{+10}~{\rm M_{\odot}}$. This limit was also reported by \cite{Ishigaki2018} who investigated the chemical enrichment of metal-poor halo stars. Our results tentatively support the work of \citet{Sukhbold2016, Sukhbold2020}. These authors found that, in models where the explosion is powered by a neutrino wind, only a fraction of the stars with masses above $20~{\rm M_{\odot}}$ successfully launch a SN explosion. At larger masses progressively more stars collapse directly onto black holes. This scenario can be tested observationally by searching for `disappearing' stars in multi-epoch imaging data \citep{Kochanek2008, Gerke2015, Reynolds2015}.

This $M_{\rm max}$ constraint that we infer here is also consistent with the black hole masses that are being detected through gravitational waves with the Laser Interferometer Gravitational-Wave Observatory \citep[LIGO;][]{Ligo2016}; the initial black hole masses detected through this experiment are often between $M\sim20-40~{\rm M_{\odot}}$. A proposed explanation of the LIGO BH masses is that they are the remnants of massive metal-poor stars. Gaia has recently detected a $M \simeq 33~{\rm M_{\odot}}$ black hole known as BH3 that is in a binary with a stellar companion \citep{Gaia2024}. This is the most massive stellar-origin black hole detected in the galaxy so far. The stellar companion of BH3 is known to be very metal-poor ([Fe/H]~$=-2.56 \pm 0.11$). Our results are in line with these observational investigations. We are finding a similar maximum mass of the enriching stellar population of some of the least polluted gas in the Universe. The lack of evidence of the chemical products of higher mass stars, may suggest that stars above this mass range are not enriching their environment and, perhaps instead, they are directly collapsing to black holes. 

\subsubsection{Model sensitivity}
This model currently only accounts for the yields from CCSNe (not those of lower mass stars). A consequence of this is that we are only sensitive to the IMF of these relatively high mass stars and not the lower mass end of the IMF. The most sensitive tests of the low mass end of the Pop. III IMF stem from empirical observations of nearby stellar populations \citep{Hartwig2015, Rossi2021}. Furthermore, the tentative redshift evolution shown in Figure~\ref{fig:co} may suggest that the yields of lower mass stars are influencing the [C/O] abundances of these EMP DLAs. Thus, the assumption of enrichment solely by CCSNe may not be the best choice. We reserve the inclusion of the yields of lower mass stars in the chemical evolution models for future work. For now, we rerun the analysis excluding [C/O] to mitigate the impact of potential contamination. The results of the enrichment model are stable against these changes. The slope of the IMF is less well-constrained (as is the explosion energy) but the inferences on $N_{\star}$ and M$_{\rm max}$ are virtually unchanged. 

\subsubsection{Reproducing the measured abundances}
Using our stochastic enrichment model, we can generate the expected abundance pattern given the inferred parameter distributions. Figure~\ref{fig:new_data_w_popiii_model} shows the observed abundances of [C/O] vs [Si/O] and [Al/O] vs [Fe/O] alongside the abundance distributions expected from this model. While the observed [C/O] and [Si/O] abundances are well-recovered, it appears more challenging to reproduce the [Al/O] and [Fe/O] abundances of these EMP DLAs. This highlights both the importance of considering multiple relative abundances simultaneously and the sensitive nature of abundance pattern fitting. Indeed, it is worth noting that the [Al/O] abundances of the \citet{Cayrel2004} stellar sample are markedly subsolar, with typical values of [Al/O]~$\sim-1.16$. By construction, these stellar data are well-modelled by the fiducial \citetalias{HegerWoosley2010} yields (see Figure 12 of \citetalias{HegerWoosley2010}). Thus, it may be impossible to reproduce the two EMP DLAs in this sample with [Al/O] abundances that are close to the solar value. Since these data are best modelled by a low number of enriching stars, we will now consider the possibility that these DLAs have been enriched by a single Population III SN. \\

\subsubsection{Enrichment by one Pop. III SN}

\begin{figure}
    \centering
    \includegraphics[width=\columnwidth]{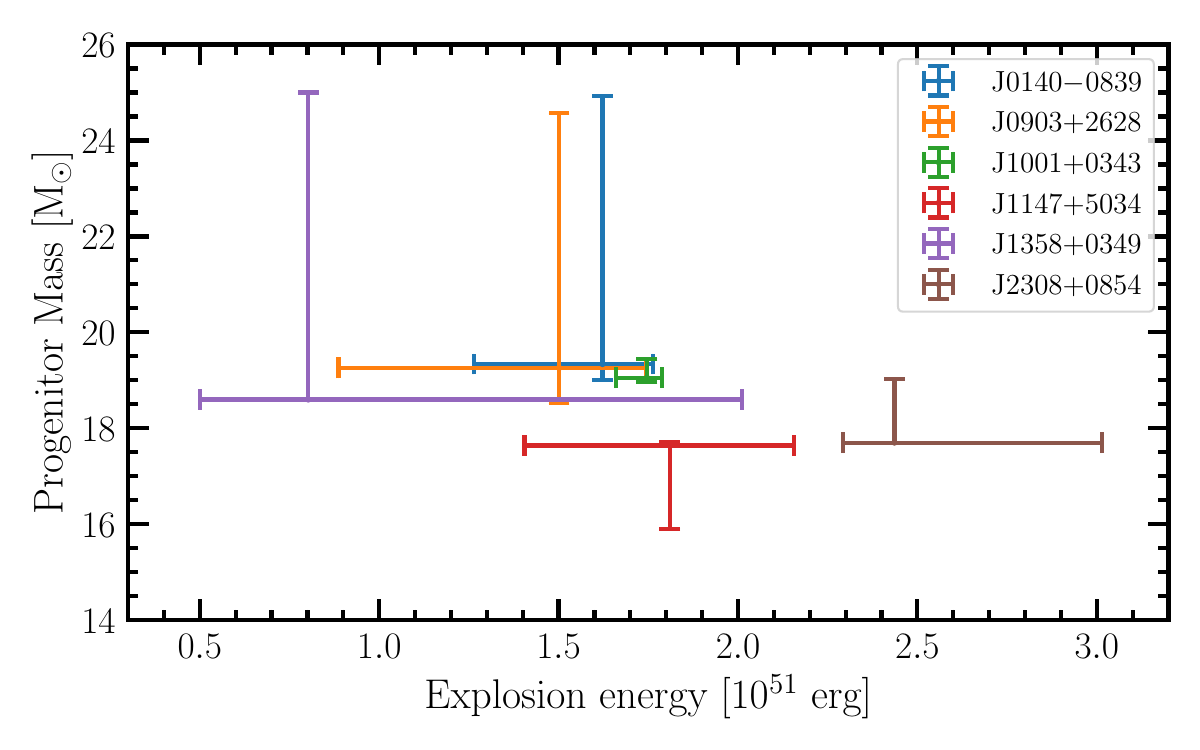}
    \caption{The inferred progenitor masses as a function of their associated explosion energies from our investigation of the properties of the individual stars that can accurately model the abundances of each EMP DLA in our sample. The properties of the progenitor associated with each DLA have been found independently from one another. We represent the resulting parameter distributions by their median value and their interquartile range. These distributions are mostly non-gaussian. In some cases, the median is close to an edge of the region bounded by the interquartile range. This is caused by a strong peak in the distribution around a particular parameter value. Compared to the full parameter space that has been explored, these DLAs show remarkably consistent progenitor properties.
    }
    \label{fig:mcmc_1N}
\end{figure}

Given that the inferred number of enriching stars is consistent with enrichment from a single Pop. III SN, we explore the properties of the individual stars that may have enriched each EMP DLA in our sample. For the first time, we investigate the distribution of recovered Pop. III SNe parameters across the EMP DLA population. The results of considering a single Pop III progenitor for each EMP DLA in turn are shown in Figure~\ref{fig:mcmc_1N}. In this scenario, we allow the degree of mixing to vary as a free parameter with a uniform prior. We show the distribution of the progenitor masses as a function of the corresponding explosion energy. The parameters and their associated errors are given by the median and interquartile range of the associated posterior distributions. Each DLA is found to be best modelled with a relatively low mass progenitor with $M \simeq 18~{\rm M_{\odot}}$. The explosion energies span a similar range to that found in our earlier analysis. In all cases, these EMP DLAs are best modelled by a progenitor that underwent a SN with $E_{\rm exp}<3 \times10^{51}$~erg. Given the extent of the explored parameter space, the best-fit progenitor properties show minimal scatter across this sample of EMP DLAs. This may suggest that these DLAs share a relatively consistent enrichment history. 

\subsubsection{Pop. III vs. Pop. II enrichment}
At this time, our model is unable to distinguish between the enrichment between one Population III star and one Population II star with the same properties. The adopted nucleosynthetic models indicate that the yields of Pop. III and Pop. II stars are extremely similar for the abundance ratios considered in this work. The $Z=0$ \citetalias{HegerWoosley2010} yields adopted are broadly consistent with the $Z = 1/1000~{\rm Z_{\odot}}$ yields presented in \citet{WoosleyWeaver1995}. Thus, a caveat to this analysis is that these DLAs may be equally well-modelled by two Population II stars rather than two Population III stars. Indeed, there is also the possibility of enrichment by a low number of both Pop. II and Pop. III stars. We adopt the \citetalias{HegerWoosley2010} yields given the extensive parameter space that is explored. The key result of the analysis is the low number of enriching stars that has been inferred; it is more difficult to reproduce such a low number of enriching stars with predominantly Pop. II stars. Given the low value of $N_{\star}$, we therefore consider it more likely that these environments are predominantly enriched by Pop. III stars.

\subsection{DLA homogeneity?}
\label{sec:homog}
Here we focused on the chemical abundances of a \emph{sample} of neutral gas reservoirs at cosmic noon. In particular, we have looked at the intrinsic scatter present in the current data. To confidently determine the source of this scatter, one must also be confident that the sample is composed of the same class of objects. 

When restricted to single sightline investigations, this can be challenging to discern. The hosts of DLAs may be a `mixed bag' of galaxies with different morphologies and associations; from protoclusters to the circumgalactic medium (CGM) \citep{Pettini1990, Haehnelt1998, Pontzen2008, Rahmati2014}. Indeed, recent simulations have highlighted that spatially distinct gas can appear as a single DLA or metal component when observed using a single sightline \citep{Mandelker2021, Marra2024}. Within this framework, it is interesting to consider whether there is an inherent difference between the kinematically simple gas clouds that have been discovered here compared to the more complex structures typically observed. The environments of these absorbers may also provide insight into the origins of the kinematics of the observed gas clouds \citep{Mackenzie2019, Lofthouse2022}. 

Simulations of the most metal-poor DLAs suggest that they may be the descendants of halos with virial masses of $M=10^{7}~{\rm M}_{\odot}$ \citep{Salvadori2012}. Over the course of their evolution, they are associated with low star formation rates \citep{Yuan2016}. Indeed, it has been suggested that the most metal-poor DLAs found at $z\sim3$ may represent gas that has gone on to form the stars that are observed within the local UFD galaxy population \citep{Webster2015b}. This has further been supported with chemical evolution models that infer the physical properties of these most metal poor DLAs (e.g., the associated volume density, neutral gas fraction, stellar and gaseous mass content; \citealt{CookePettiniJorgenson2015}; \citealt{Welsh2019}). These properties have been found to be consistent with the equivalent properties within UFDs \citep{MartinDeJongRix2008}. Though, there are also alternative simulations that disfavour this scenario \citep[e.g.][]{Jeon2019}. A future avenue of investigation could be to investigate whether this discrepancy can be resolved when accounting for the small scale kinematic structures of these most metal-poor DLAs (both observationally and within cosmological simulations). \\

\section{Conclusions}
\label{sec:conc}
We present new data of 8 near-pristine absorption line systems discovered at cosmic noon and use these data to investigate early chemical enrichment. This sample includes all currently known EMP DLAs and marks the first analysis of 4 new metal-poor systems. We pay particular attention to the behaviour of [O/Fe] in the EMP regime. Our main conclusions are as follows:
\begin{enumerate}
    \item  We have presented new data of the previously known EMP DLAs J0140$-$0838 and J1358$+$0349 (Figures~\ref{fig:j0140} and \ref{fig:j1358}). These new data improved the precision of the [Fe/H] and [O/Fe] determinations by a factor of 3. In both cases, the latest abundances confirm that these are two of the most Fe-poor DLAs discovered to date. The [O/Fe] abundances are moderately elevated and consistent with the [O/Fe] abundances indicated by previous determinations. The improved precision is thanks to our broader wavelength coverage that includes multiple Fe\,{\sc ii} transitions covering a range of oscillator strengths.
    \item We report the discovery of two new VMP DLAs and two new EMP DLAs (Figures~\ref{fig:j0239} - \ref{fig:j2308}). With these data, combined with our recent analysis of the most metal-poor DLA currently known \citep[][]{Welsh2023}, our survey has doubled the number of known EMP systems.
    
    \item With these new statistics we re-evaluate the behaviour of [O/Fe] as a function of the Fe-metallicity (Section~\ref{sec:chem} and Figure~\ref{fig:ofe_new}). We find further evidence of enhanced scatter of the [O/Fe] abundances at the lowest metallicities (Figure~\ref{fig:intsca}). The distinction between the typical [O/Fe] of EMP DLAs compared to VMP DLAs is less severe than previously indicated. We quantify these results using the intrinsic scatter of [O/Fe] for VMP and EMP DLAs. We find that EMP DLAs are well-modelled with [O/Fe]$_{\rm cen} =+0.50 \pm 0.04$ and an intrinsic scatter of $\sigma_{\rm int \, ,\, [O/Fe]} = 0.13_{-0.04}^{+0.06}$. Both parameters are higher than that found for VMP DLAs, with [O/Fe]$_{\rm cen} =+0.40 \pm 0.02$ and $\sigma_{\rm int \, ,\, [O/Fe]} = 0.06\pm 0.02$.  
    
    \item We exploit the detection of both Si\,{\sc ii} and Si\,{\sc iii} absorption features in the spectrum towards J1147$+$5034, in combination with photoionization models, to investigate the potential need for ionization corrections (Section~\ref{sec:ioncorr}). We find that in all cases the potential corrections are $<0.15$~dex (Figure~\ref{fig:ic}). These corrections could only serve to further increase the [O/Fe] values of the EMP DLAs discovered with this programme.
    
    \item We also present an updated analysis of the evolution of [C/O] as a function of redshift (Figure~\ref{fig:co}). We find that the most metal-poor systems (i.e., those within 0.1~dex of the EMP regime) show a gradually increasing [C/O] ratio at lower redshifts. This may be a sign of the onset of enrichment from the first low and intermediate mass stars associated with these gas clouds.
    
    \item We employ a stochastic chemical enrichment model to investigate the potential enrichment histories of these EMP gas clouds (Section~\ref{sec:model}). We find that the [C/O], [Al/O], [Si/O], and [Fe/O] abundances of these DLAs are best modelled with an IMF slope that is consistent with a Salpeter distribution with a slight preference towards a more top-heavy IMF (Figures~\ref{fig:mcmc_res} and \ref{fig:new_data_w_popiii_model}). The number of enriching stars is found to be $N_{\star} =2\pm1$. This is consistent with the number of enriching stars that may be expected from one Pop. III minihalo \citep{Skinner2020}. The maximum mass of the enriching stars is $M_{\rm max}=32_{-4}^{+10}~{\rm M_{\odot}}$. Interestingly, this is broadly consistent with the black hole masses that have been detected through gravitational waves \citep{Ligo2016} and is similarly in-line with the most massive stellar origin galactic black hole (BH3) that has recently been detected with Gaia alongside a metal-poor companion \citep{Gaia2024}. A possible explanation of the LIGO detected black holes, that is supported by the recent Gaia detection, is that they are the remnants of massive metal-poor stars in the early Universe that failed to explode as core-collapse SNe. Indeed, we are finding a similar maximum mass of the enriching stellar population of some of the least polluted gas in the Universe. The lack of evidence for the chemical products of higher mass stars may suggest that stars above this mass range are not enriching their environment and, perhaps instead, they are directly collapsing to black holes. This limit was also reported by \citet{Ishigaki2018} who investigated the chemical enrichment of metal-poor halo stars.
    
    \item We propose that an interesting avenue of future investigation is whether there is an intrinsic difference between kinematically simple EMP DLAs compared to those with more complex structures (Section~\ref{sec:homog}). Indeed, this could be an avenue to discern the range of potential hosts of these most metal-poor gas reservoirs \citep{Salvadori2012, Webster2015b, Jeon2019, Mandelker2021}.
    
\end{enumerate} 
 This survey has enabled the first detailed understanding of EMP DLAs, and the potential of these gas clouds to study the properties of the first stars. Upcoming surveys \citep[e.g. DESI, WEAVE, and 4MOST][]{Dalton2012, deJong2012, DESI2016, DESI2016b, DESI2023, Pieri2016} will provide new EMP DLA candidates to study the chemical enrichment of near-pristine gas. If an elevated [O/Fe] abundance can be attributed to enrichment by metal-free stars, then the DLAs analysed in this work may provide a signpost to some of the most pristine environments in the high redshift universe, and would be an ideal place to search for Population III host galaxies and perhaps even the light from Population III SNe using the \emph{James Webb Space Telescope}.

\begin{acknowledgements}
      We thank the anonymous referee for their prompt and thorough review of this paper. We thank Valentina D'Odorico, Paolo Molaro, and Trystyn Berg for helpful comments and feedback. This paper is based on observations collected at the William Herschel Telescope (WHT) at the Roque de los Muchachos Observatory on the island of La Palma in the Canary Islands, Spain (programme ID:  W/2019B/04). L.~A.~W  is particularly grateful to Lucia Suarez-Andres and all of the staff astronomers at the WHT who provided significant support during the observing runs throughout the months of Sep - Dec 2019. This paper is further based on observations collected at the European Organisation for Astronomical Research in the Southern Hemisphere, Chile, and at the W. M. Keck Observatory which is operated as a scientific partnership among the California Institute of Technology, the University of California and the National Aeronautics and Space Administration (programme IDs in Table 1). The Observatory was made possible by the generous financial support of the W. M. Keck Foundation. The authors wish to recognize and acknowledge the very significant cultural role and reverence that the summit of Maunakea has always had within the indigenous Hawaiian community. We are most fortunate to have the opportunity to conduct observations from this mountain. We are also grateful to the staff astronomers at the VLT and Keck Observatory for their assistance with the observations. 
This work has been supported by Fondazione Cariplo, grant No 2018-2329. During this work, R.~J.~C. was supported by a Royal Society University Research Fellowship. We acknowledge support from STFC (ST/L00075X/1, ST/P000541/1). This project has received funding from the European Research Council (ERC) under the European Union's Horizon 2020 research and innovation programme (grant agreement No 757535). This work used the DiRAC@Durham facility managed by the Institute for Computational Cosmology on behalf of the STFC DiRAC HPC Facility (www.dirac.ac.uk). The equipment was funded by BEIS capital funding via STFC capital grants ST/P002293/1, ST/R002371/1 and ST/S002502/1, Durham University and STFC operations grant ST/R000832/1. DiRAC is part of the National e-Infrastructure.
This research has made use of NASA's Astrophysics Data System.
\end{acknowledgements}


%
%
\bibliography{references}
\bibliographystyle{aa}

\begin{appendix}

\section{Telluric line removal}
\label{appen:skylines}

\begin{figure*}
    \centering
    \includegraphics[width=0.8\textwidth]{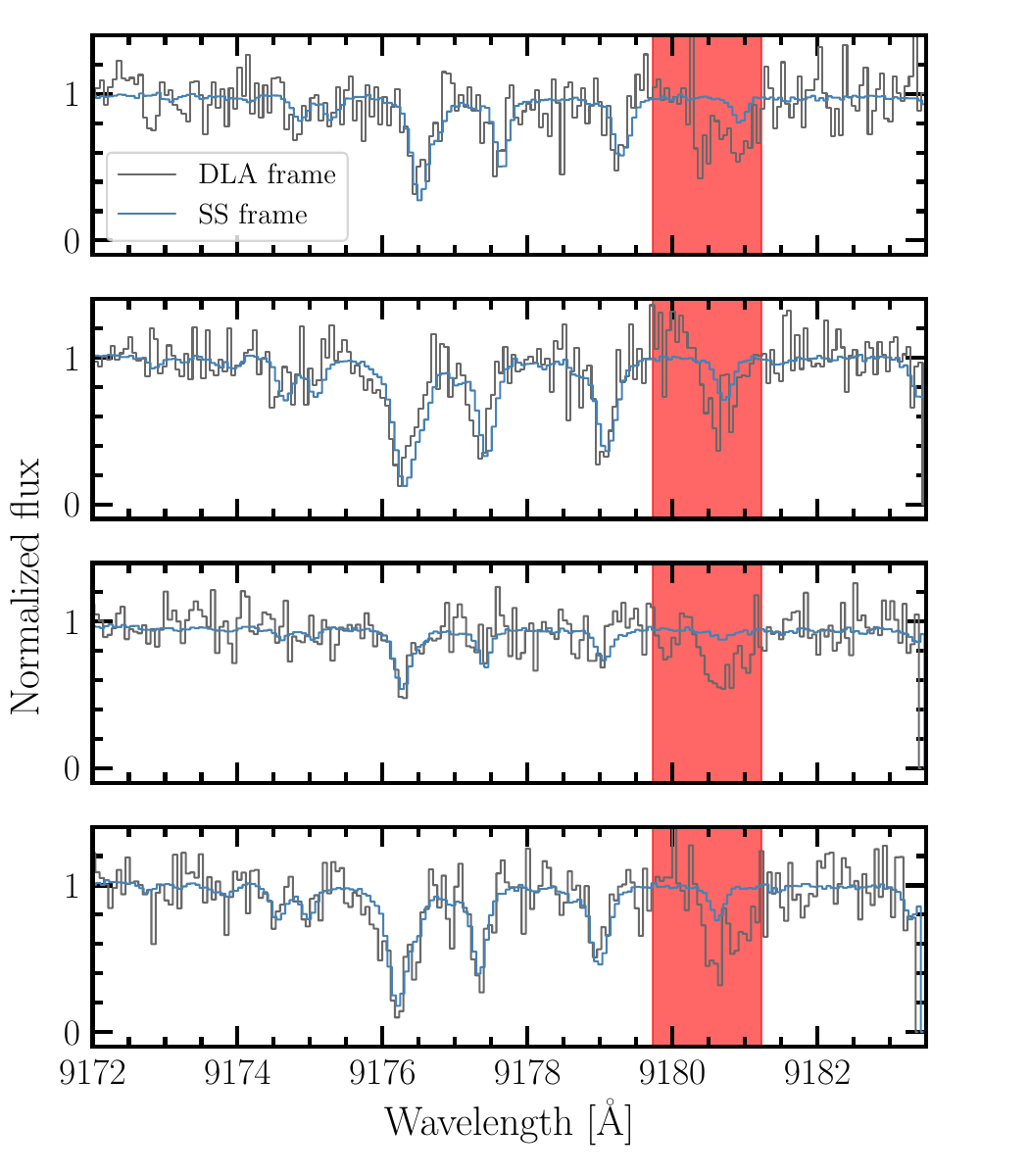}
    \caption{A subset of the UVES data taken for J1358$+$0349 (grey histogram) overplotted with the associated observations of the telluric standard star (blue histogram). These panels show the first four exposures of this target. The region shaded in red covers the 1.5~\AA\ region of the spectrum centred on where the \fe{2382} feature of the central DLA component should fall.}
    \label{fig:J1358_tell}
\end{figure*}

In this appendix we show the treatment of telluric absorption removal for the spectrum of J1358$+$0349. Prior to combining the exposures, we first remove the telluric absorption near the Fe\,{\sc ii} features of interest using a telluric standard star. The \fe{2382} line appears in two orders and we perform this correction to both orders separately. An example of the data near the \fe{2382} line prior to the removal of the telluric lines is shown in Figure~\ref{fig:J1358_tell}. This figure shows the relevant wavelength region for a selection of exposures alongside the equivalent data from the observations of the telluric star. \\

\end{appendix}

\end{document}